\documentclass[aip,jcp,reprint,twocolumn]{revtex4-1}
\usepackage{graphicx}
\usepackage{xcolor}

\newcommand{\beq}{\begin{equation}}
\newcommand{\eeq}{\end{equation}}
\newcommand{\bea}{\begin{eqnarray}}
\newcommand{\eea}{\end{eqnarray}}

\usepackage[colorlinks=true, linkcolor=black, urlcolor=blue, citecolor=blue]{hyperref}

\begin{document}

\title{The protein escape process at the ribosomal exit tunnel has conserved
mechanisms across the domains of life}

\author{Phuong Thuy Bui}
\affiliation{Graduate University of Science and Technology, Vietnam Academy of
Science and Technology, 18 Hoang Quoc Viet, Cau Giay, Hanoi 11307, Vietnam}
\affiliation{Institute of Theoretical and Applied Research, Duy Tan University,
Hanoi, 100000, Vietnam}
\affiliation{Faculty of Pharmacy, Duy Tan University, Da Nang, 550000, Vietnam}

\author{Trinh Xuan Hoang}
\email[Corresponding author, E-mail: ]{txhoang@iop.vast.vn}
\affiliation{Institute of Physics, Vietnam Academy of Science and Technology,
10 Dao Tan, Ba Dinh, Hanoi 11108, Vietnam}
\affiliation{Graduate University of Science and Technology, Vietnam Academy of
Science and Technology, 18 Hoang Quoc Viet, Cau Giay, Hanoi 11307, Vietnam}

\begin{abstract}
The ribosomal exit tunnel is the primary structure affecting the release of
nascent proteins at the ribosome. The ribosomal exit tunnels from different
species have elements of conservation and differentiation in structural and
physico-chemical properties. In this study, by simulating the elongation and
escape processes of nascent proteins at the ribosomal exit tunnels of four
different organisms, we show that the escape process has conserved mechanisms
across the domains of life. Specifically, it is found that the escape process
of proteins follows the diffusion mechanism given by a simple diffusion model
and the median escape time positively correlates with the number of hydrophobic
residues and the net charge of a protein for all the exit tunnels considered.
These properties hold for twelve
distinct proteins considered in two slightly different and improved G\=o-like
models.  It is also found that the differences in physico-chemical properties
of the tunnels lead to quantitative differences in the protein escape times. In
particular, the relatively strong hydrophobicity of the E. coli's tunnel and
the unusually high number of negatively charged amino acids on the tunnel's
surface of H. marismortui lead to substantially slower escapes of proteins at
these tunnels than at those of S. cerevisisae and H. sapiens.
\end{abstract}

\maketitle

\section{Introduction}

The ribosomal exit tunnel is a narrow structure connecting
the peptidyl transferase center (PTC), where the polypeptide polymerization
takes place during translation, to the surface of the ribosome. It is the
first structure encountered by the nascent polypeptides and is the only passage
for nascent proteins to be released from the ribosome. The ribosomal exit
tunnel is believed to play important roles in translation regulation
\cite{Ito2002,Ito2014,Wilson2016} as well as 
co-translational protein folding \cite{Cabrita2010,Cavagnero2011,Rodnina2017}. 
The tunnel dimensions, 10--20~{\AA} in width and 80--100~{\AA} in length,
allow it to accommodate up to $\sim$40 amino acids \cite{Voss2006} 
but limit the size of the folded peptide inside the tunnel
\cite{Deutsch2009}. In general, the protein and RNA composition of the ribosome
can vary in different domains and different species, leading to different
structural details of the exit tunnel. A comparison of the tunnel structures
from a range of species have shown certain similarities and differences
\cite{Khanh2019}. For example, it has been shown that the upper part of the
tunnel, near the PTC, is relatively conserved across the species. On the other
hand, the lower part of the tunnel is substantially narrower in eukaryotes than
in bacteria, which may have implications for antibiotic resistance
\cite{Kannan2011}.

The post-translational escape of nascent proteins at the ribosomal
exit tunnel is the final release of a protein from the ribosome when the
protein's N-terminus is no longer attached to the PTC. This process is a
necessary step of a nascent protein to empty the ribosomal exit tunnel
for the next translation process and to complete its own folding to the native
state. Only very few studies have addressed this process until recently.
In earlier works \cite{Thuy2016, Thuy2018}, by coarse-grained simulations in
the G\=o-like models, we have shown that the escape process is assisted by the
folding of the nascent protein and is akin to the diffusion of a Brownian
particle in a linear potential field. In more recent works, by using the
atomistic tunnel of H. marismortui, it was shown that the roughness of the exit
tunnel can increase the difficulty of nascent proteins to escape
\cite{Thuy2020} and that the escape time is modulated by energetic interactions
of the protein with the exit tunnel, such as hydrophobic and
electrostatic interactions \cite{Thuy2021}. Another study with the E. coli's
tunnel suggests that electrostatic interaction can extremely delay
the protein escape \cite{Nissley2020}.

The present study is aimed to extend our understanding of the protein escape
process at the ribosomal exit tunnels of different species. In particular, we 
consider the exit tunnels from four organisms, namely E. coli, H.
marismortui, S. cerevisiae and H. sapiens, which are representatives
from all three domains of life (bacteria, archaea and eukarya). Their
ribosome structures have been experimentally determined at high resolutions
allowing us to have atomic details for the tunnel models used in the simulations. 
The belief is that the differences in the structural and chemical details of
the exit tunnels considered will help us to have a more complete picture of the
protein escape process at ribosomal exit tunnels.

We used the same simulation approach as in the previous study \cite{Thuy2021} to
study the escape process but with a larger set of proteins and with 
improved models for the nascent proteins. The G\=o-like models in the
present study, namely the G\=o-MJ and G\=o-MJ-nn models, incorporate the
well-known Miyazawa-Jernigan's contact energy matrix in the depths of
the Lennard-Jones potentials for native and non-native contacts, thereby
to a certain degree take into account the effects of the amino acid sequences
in the escape and folding of these proteins. The energy parameters in the
G\=o-like models are also rescaled such that the melting temperature in
the model matches the experimental melting temperature of each protein.

We will show that while there are significant variations in the escape times
among the exit tunnels of different organisms, the mechanisms governing
the protein escape are remarkably similar at different exit tunnels
suggesting that they are conserved across the domains of life.

\section{Models and Method}

\subsection{Improved G\=o-like models}

G\=o-like models have been widely used to study the protein folding dynamics
due to their simplicity and effectiveness \cite{Baker2000,Hills2009,Takada2019}. 
They are a class of models which emphasizes the importance of native
interactions \cite{Go1975} and can be applied to any protein with a known
native structure. 
In this work, we used two variants of improved G\=o-like models to simulate
nascent proteins: the first one incorporates variable strengths of the
potentials
for native contacts, and the second one includes also
attractive potentials for non-native contacts.
These models partially take into account the effects of the amino acid sequence
through the use of the Miyazawa-Jernigan matrix for inter-residue contact
energies \cite{MJ1996}, in a similar manner to 
other G\=o-like models used in the literature
\cite{Brooks2003,Best2005,Takada2015,Quyen2021}.

\subsection*{G\=o-MJ model}

The G\=o-MJ model is modified from the one of Clementi {\it et al.}
\cite{Clementi2000} by adding a variation in the strengths of the potentials
for native contacts.  Considering only the C$_\alpha$ atoms, the potential
energy of a protein in a given conformation is given by
\begin{eqnarray}
V_{\text{G\=o-MJ}} &=& \sum_\mathrm{bonds} K_b (r_{i,i+1} - r^*_{i,i+1})^2 
+ \sum_\mathrm{angles} K_\theta (\theta - \theta^*)^2 \nonumber \\
&+& \sum_\mathrm{dihedrals} \sum_{n=1,3} K_\phi^{(n)} [1 - \cos(n(\phi - \phi^*))] 
\nonumber \\ 
&+& \sum_{j>i+3}^{\text{native}} \epsilon^{\mathrm{NC}}_{ij} \left[
5 \left(\frac{r^*_{ij}}{r_{ij}}\right)^{12} -
6 \left(\frac{r^*_{ij}}{r_{ij}}\right)^{10} 
\right] \nonumber \\
&+& \sum_{j>i+3}^{\text{non-native}} 
\epsilon \left(\frac{\sigma}{r_{ij}}\right)^{12} ,
\end{eqnarray}
where the terms on the right side correspond to the potentials on the bond
lengths, bond angles, dihedral angles, native contacts and non-native contacts,
respectively, as described in detailed elsewhere~\cite{Clementi2000,Thuy2020}.
The native contacts are determined from an all-atom consideration
\cite{Cieplak2002} of the protein structure from the Protein Data Bank (PDB)
and the atomic van der Waals radii \cite{Tsai1999}.  $r_{ij}$ is the
distance between residue $i$ and residue $j$, the $^*$ symbol denotes the
native state's value, $\sigma$ is an effective diameter of amino acids,
and $\epsilon$ is an energy parameter. 
The value of $\epsilon^\mathrm{NC}_{ij}$,
which sets the potential depth for a native contact, is calculated as
\begin{equation}
\epsilon^\mathrm{NC}_{ij} =  
\frac{ n_{ij} e_\mathrm{HB} + e_{\mathrm{MJ}} (s_i,s_j) }
{u} \,\epsilon ,
\end{equation}
where $n_{ij}$ is the number of hydrogen bonds between residue $i$ and residue
$j$ in the native state, $e_\mathrm{HB} = 1.5$~kcal/mol is a hydrogen bond's
energy, $e_\mathrm{MJ}(s_i,s_j)$ is the inter-residue contact energy for the
pair of amino acids of the types $s_i$ and $s_j$ given by the Miyazawa-Jernigan
matrix \cite{MJ1996} with the energy converted to kcal/mol and given in the
absolute value; $u$ is a normalizing factor, such that the average energy of
all the native contacts is $\epsilon$. Other parameters in the model are
$\sigma=5$~{\AA}, $K_b = 100~\epsilon\,${\AA}$^{-2}$, 
$K_\theta=20~\epsilon\,$(rad)$^{-2}$,
$K_\phi^{(1)} = \epsilon$, 
$K_\phi^{(3)} = 0.5 \epsilon$. 

\subsection*{G\=o-MJ-nn model}

In the G\=o-MJ-nn model, the last term in Eq. (1) is replaced by
\begin{equation}
\sum_{j>i+3}^{\text{non-native}} 
\epsilon^{\text{NN}}_{ij}
\left[
5 \left(\frac{\sigma_1}{r_{ij}}\right)^{12} -
6 \left(\frac{\sigma_1}{r_{ij}}\right)^{10} 
\right] ,
\label{eq:gomjnn}
\end{equation}
which provides attraction to the non-native contacts.
The potential depth for a non-native contact is calculated as
\begin{equation}
\epsilon^{\text{NN}}_{ij} = f \,
\frac{ e_\mathrm{MJ}(s_i,s_j) }
{u} \,\epsilon ,
\end{equation}
where $f$ is a factor that sets the relative strengths of non-native
contacts. In the present study, we used
$\sigma_1=5.5$~{\AA} and $f=0.4$.

In both the G\=o-MJ and G\=o-MJ-nn models, $\epsilon$ is the single parameter
that sets the energy scale of the whole protein. Because temperature effects
are important for the dynamics of proteins, especially for their diffusion in
the ribosomal tunnel, it is important to have the correct energy scale for each
protein. Following the previous work \cite{Thuy2021}, we determined $\epsilon$
individually for each protein by fitting the melting temperature
in the model to the experimental melting temperature, $T_m$. 
The melting temperature in the model is defined by $T_\mathrm{max}$,
the temperature of the specific heat's maximum of a protein obtained
by simulations. The parameter $\epsilon$ is calculated as
$
\epsilon = \frac{(273+T_m)}{503.2195 \times T_\mathrm{max}} \ \text{(kcal/mol)} ,
$
where $T_\mathrm{max}$ is given in units of $\epsilon/k_B$ and $T_m$ is given in $^\circ$C.
The values of $T_\mathrm{max}$ and $\epsilon$ in both G\=o-MJ and G\=o-MJ-nn
models for a list of 12 proteins considered are given in Table~S1.

\subsection{Tunnel model}

Models of the exit tunnels are constructed based on the PDB structures of
the large ribosomal subunits of the organisms. The structures of the PDB IDs
7k00 \cite{Watson2020}, 1jj2 \cite{Steitz2005}, 5gak \cite{Schmidt2016} and
4ug0 \cite{Khatter2015} and considered for the ribosomes of E. coli, H.
marismortui, S. cerevisiae
and H. sapiens, respectively.  The model considers all the heavy atoms for
ribosomal RNA but only C$_\alpha$'s for ribosomal proteins.  To reduce
computational time, we kept only atoms within a cylinder of radius $R$ centered
around an approximate chosen tunnel axis for the tunnel model. The value of $R$
must be sufficiently large to enclose the atoms of the tunnel's wall. We have
chosen $R=30$~{\AA} for the ribosome tunnels of H. marismortui, S.  cerevisiae
and H. sapiens and $R=45$~{\AA} for that of E. coli. The model also ignores the
motion of the ribosome, thus all the tunnel atoms are kept fixed during the
simulations. 

For interactions of the tunnel with nascent proteins, the model used in this
study is the T3 model described in Ref.~\cite{Thuy2021}, which contains three
types of interactions: excluded volume, hydrophobic and electrostatic. Details
of the interaction potentials are given in Ref.~\cite{Thuy2021}.  In short, the
exclude volume interaction provides a short-range repulsion between the
tunnel's atoms and the nascent chain's residues. The hydrophobic interaction
gives an attraction between hydrophobic residues (Ile, Leu, Phe, Met, Val, Pro,
Trp) of a nascent protein and those of the same type in ribosomal proteins via
a 10-12 Lennard-Jones potential. The depth of this potential is constant for
all pairs of hydrophobic residues and is equal to $\epsilon_\mathrm{hydr} =
1.2$~kcal/mol.  The electrostatic interaction is given by a screened Coulomb
potential from the Debye-H\"uckel theory with the Debye's screening length 
$\lambda_D=10$~{\AA}. 
The electrostatic interaction is considered between all charged residues
of a nascent protein and all charged centers of rRNA and ribosomal proteins. 
In rRNA, each phosphorus atom is assigned with the charge $q=-1e$.
In nascent and ribosomal proteins, lysine and arginine are given 
with the charge $q=+1e$, whereas aspartic acid and glutamic acid are given with
$q=-1e$. The charges of amino acids are assumed to be concentrated on the
C$_\alpha$ atoms.

\subsection{Simulation method}

A molecular dynamics (MD) method based on the Langevin equation of motion is
used to simulate the motions of nascent chains. Details of the method are given
in Ref.~\cite{Thuy2016}. We adopt a reduced unit system such that the mass unit
is the average mass $m$ of amino acids, the length unit is the effective
diameter $\sigma$ of amino acids, and the energy unit is kcal/mol. The friction
coefficient of amino acids used in the simulations is $\zeta = 1\,\sqrt{m
\sigma^{-2}(\mathrm{kcal/mol})}$. Given that $m=120$~g/mol and
$\sigma=5$~{\AA}, the simulation time is measured in the units of
$\tau= \sqrt{m \sigma^2/\mathrm{(kcal/mol)}} \approx 3$~ps.  This value
of the time unit, suitable for the low-friction regime \cite{Veitshans1997},
results in a much shorter timescale of the simulation folding times than the
real folding times. It has been shown that the correct timescale can be reached
by
simulations by increasing $\zeta$ to its realistic value and using the
high-friction estimate, $\tau_H = 3$~ns, of the time unit
\cite{Veitshans1997,Klimov1997,Thuy2021}.

For an isolated protein, the temperature of the specific heat's maximum
$T_\mathrm{max}$ is determined from the temperature dependence of the specific
heat by using replica-exchange
molecular dynamics (REMD) simulations \cite{Sugita1999} and the weighted
histogram analysis method
\cite{Ferrenberg1989,Kumar1992}.
For studying the protein escape at a ribosome tunnel,
both the translation process and the escape process are simulated.
In the translation process, a nascent chain is elongated at
the position of the PTC at a constant rate corresponding to a growth time $t_g$
per residue. $t_g$ must be chosen sufficiently large such
that the escape properties are converged given that the growth times in
cells are orders of magnitude larger 
than in simulations. We used $t_g=400\tau$ for most proteins, and
$t_g=2000\tau$ for proteins that are kinetically trapped at the tunnel.
The escape time is measured from the moment of complete elongation (the
C-terminal residue is released from the PTC) until the nascent protein has
fully escaped the tunnel. All simulations of the translation and
escape processes of proteins are carried out at the room temperature $T=300$~K. 
Typically, the escape time distribution and the escape probability are
calculated from 1000 independent trajectories for each protein.

\subsection{Diffusion model}

The diffusion model \cite{Thuy2018} considers the protein escape process as the
diffusion of a Brownian particle in a one-dimensional potential field $U(x)$
with $x$ the position of the particle.  Such process is governed by the
Smoluchowski equation. Given the linear form $U(x)=-kx$ of the external
potential, where $k$ is a
constant force acting on the particle, the distribution of the escape time can
be obtained from an exact solution of the Smoluchowski equation and is given by
\cite{Thuy2018}
\begin{equation}
g(t)=\frac{L}{\sqrt{4\pi D t^3}} \exp\left[
-\frac{(L - D\beta k t)^2}{4Dt}
\right] ,
\label{eq:gt}
\end{equation}
where $L$ is diffusion distance equal to the tunnel length, $D$ is the
diffusion constant assumed to be position independent, $\beta=(k_B T)^{-1}$ is
the inverse temperature where $k_B$ is the Boltzmann constant.
Interestingly, the escape time distribution in Eq. (\ref{eq:gt}) 
can fit the data from various simulations of protein escape
in the G\=o-like model \cite{Thuy2018,Thuy2020,Thuy2021}. It has been shown
that the free energy of a protein at the ribosome tunnel is approximately
linear along an escape coordinate \cite{Thuy2016,Thuy2021},
which justifies the linear form of $U(x)$ in the diffusion model.

The distribution in Eq. (\ref{eq:gt}) gives the mean value
$\mu_t=L/(D\beta k)$ and the standard deviation 
$\sigma_t=\frac{\sqrt{2L}}{D(\beta k)^{3/2}}$ for the escape
time\cite{Thuy2018}. Note that both $\mu_t$ and $\sigma_t$ diverges when $k=0$,
for which $g(t)$ becomes a
heavy-tailed
L\'evy distribution.

\section{Results}

\subsection{Differences in physico-chemical properties of  
nascent proteins and the ribosomal exit tunnels}

This study considered twelve small globular proteins with known melting
temperatures $T_m$. They consist of
the B1 domain of protein G (1pga) \cite{Alexander1992}, 
Rop protein (1rop) \cite{Predki},
the SH3 domain (1shg) \cite{Chen1996}, 
the Z domain of Staphylococcal protein A (2spz) \cite{Myrham2020}, 
Cro repressor (1orc) \cite{Gitelson},
chymotrypsin inhibitor 2 (2ci2) \cite{Jackson1991}, 
antifreeze protein (1msi) \cite{Arribas},
cold-shock protein (1csp) \cite{Welte2020}, 
ubiquitin (1ubq) \cite{Morimoto}, 
histidine-containing phosphocarrier protein HPr (1poh) \cite{Thapar}, 
hyperthermophilic archaeal DNA-binding protein Sso10b2 (1udv) \cite{Biyani},
and barnase (1a2p) \cite{Bye2016} with the PDB IDs of their
native structures enclosed in the parentheses. The references
associated with the proteins
correspond to the experimental studies in which $T_m$ has been reported
(see Table S1 for the values of 
$T_m$ and other properties of the proteins).
For convenience, we will call the proteins by their PDB IDs.
These proteins have lengths between 56 and 108 amino acids and
distinct native structures with two all-$\alpha$, two all-$\beta$ and
eight $\alpha/\beta$ proteins. Their $T_m$ values range from
43.8$^\circ$C (for 1csp) to 157.5$^\circ$C (for 1udv).  Our analyses show
that the fraction of hydrophobic residues in their amino acid sequences
varies from $\sim$21\% to $\sim$47\%,
the fraction of positively charged amino acids varies between
$\sim$6\% and $\sim$18\%, 
and the fraction of negatively charged amino acids ranges
from $\sim$6\% and $\sim$19.6\%.
The protein net charges are from $-6e$ to $+3e$.
These properties indicate that the proteins considered have a wide range of
specificities leading to diverse interactions with the exit tunnel.

All proteins in our considerations are small-sized and cannot compare to
any protein length distribution in proteomes (Fig. S1). However, it can be
expected that the chain length does not impact the escape time, as shown in
one of our
previous studies \cite{Thuy2018}. The structural class, on the other hand, can
have a minor effect on the protein escape with $\alpha$-proteins escaping
somewhat more slowly than $\beta$-proteins \cite{Thuy2018}.  We have checked
that the structural class composition of our set of proteins is not too
different from that of Richardson's Top2018 high-quality protein structures
\cite{Richardson2021} with a strong dominance of $\alpha/\beta$-proteins (see
Table~S2).  Even though our protein set is quite small with only 12 proteins, its
protein sequences have similar ranges of hydrophobicity and fractions of
positively and negatively charged amino acids as found in various proteomes,
whose sequences are taken from the UniProt database \cite{uniprot2021}
(Figs.~S2 and S3). For example, the range of fraction of hydrophobic amino
acids in our 12 proteins is shared by 97\% of the protein population in the
human proteome, whereas the corresponding numbers for the ranges of fractions
of positively and negative charged amino acids are 92\% and 88\% (Fig.~S2
(a--c)). In the proteomes of S.  cerevisiae (Fig.~S2 (d--f)) and E. coli
(Fig.~S3 (a--c)), these percentages are also very high ranging from 84\% to
98\%. The proteins in the H. marismortui's proteome (Fig.~S3 (d--f)) tend to
have a lower fraction of positively charged amino acids and a higher fraction
of negatively charged amino acids than in the other organisms, resulting in
only 78\% and 68\% of the proteome sharing the ranges of these two fractions,
respectively, with the 12 proteins considered.  These statistics suggest that
the chosen proteins to a good extent reflect the variabilities of hydrophobic
and charge compositions of the proteins in the organisms considered, though
they do slightly worse for H. marismortui. They can be considered as
representative of typical globular proteins in terms of hydrophobic and charge
fractions in the amino acid sequences.

\begin{figure} 
\center
\includegraphics[width=8.5cm]{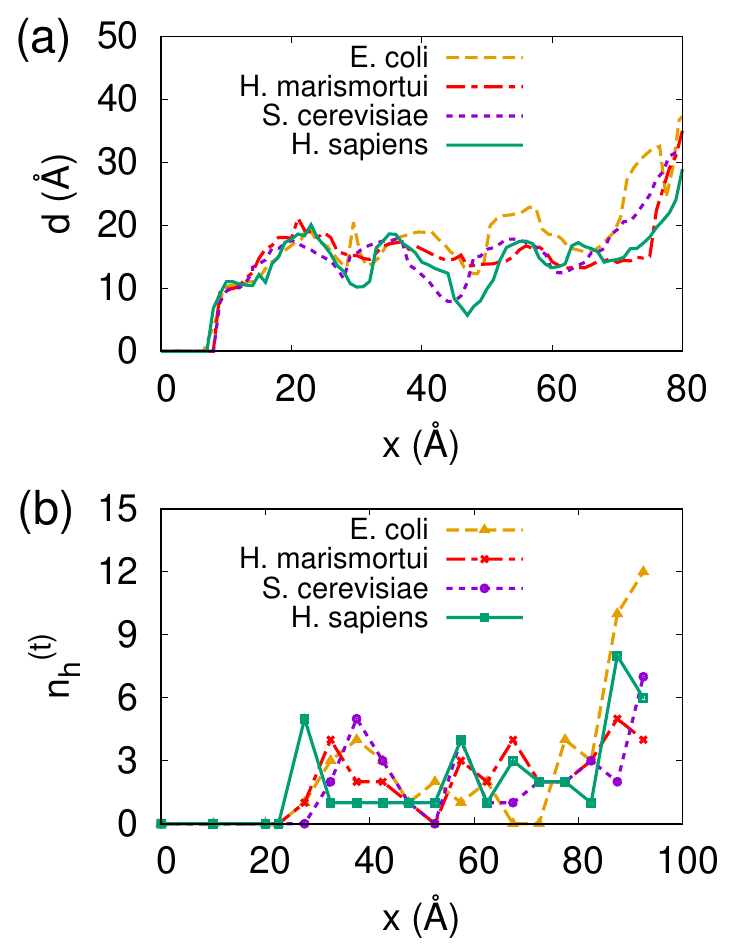}
\caption{(a) Dependence of the effective diameter, $d$, on the coordinate
$x$ along an approximate tunnel axis for the ribosomal exit tunnels of 4 
organisms considered. The lines shown are for E. coli (dashed), H. marismortui
(dash-dotted), S. cerevisiae (dotted) and H. sapiens (solid) as indicated.  
The effective diameter $d$ is calculated as $d = 2\sqrt{(S/\pi)}$, where $S$ is
the tunnel cross-sectional area accessible for a probe sphere of radius
3~{\AA}.
(b) Distribution of hydrophobic residues at the tunnel surface
along the tunnel axis for the considered species as indicated. For a given
data point at position $x$, $n_{h}^{(t)}$ is the number of surface's hydrophobic
residues within 5~{\AA} from $x$ along the tunnel axis.
}
\label{fig:4tunnels}
\end{figure}

The ribosomal exit tunnels of the four organisms considered have
notable differences and similarities. The differences in the shape of these
tunnels can be visualized through the graphs representing their effective
diameter $d$ along the tunnel axis $x$ shown in Fig.~\ref{fig:4tunnels}(a).
For each position $x$, $d$ is calculated as $d=2\sqrt{(S/\pi)}$, where $S$
is the tunnel's cross-sectional area accessible by a probe sphere of radius
3~{\AA}. Although the effective diameter does not reflect all information
about the shape of a tunnel, it already shows that the detailed shapes are
different for different species. The diameter of the H. marismortui's tunnel
appears to be the most uniform while the other tunnels show stronger variations
of $d$. The tunnel of E. coli is somewhat wider than the other tunnels
\cite{Khanh2019}.  The diameter profiles in Fig.~\ref{fig:4tunnels}(a)
also show some similarities, such as the average widths of the tunnels are
more or less the same, the tunnels become wider near the exit, and the
position at which the tunnel is narrowest appears to be about half-way from the
opening of the tunnel for all tunnels. The $d$ profile of S.
cerevisiae looks the most similar to that of H. sapiens.

\begin{table} 
\begin{tabular}{lccc}
\hline
Organism & $N_h^{(t)}$ & $N_{+}^{(t)}$ & $N_{-}^{(t)}$ \\
\hline
E. coli & 46 & 30 & 4  \\
H. marismortui & 35 & 26 & 19 \\
S. cerevisiae & 33 & 32 & 4 \\
H. sapiens & 37 & 26 & 1 \\
\hline
\end{tabular}
\caption{Hydrophobic and charged properties of ribosomal exit tunnels' 
surfaces of the organisms. For each organism, the listed properties are the
number of hydrophobic residues ($N_h^{(t)}$), the numbers of positively
($N_{+}^{(t)}$) and negatively ($N_{-}^{(t)}$) charges of ribosomal amino acid
residues that are found at the tunnel's surface. Note that these properties do
not refer to the ribosomal RNA.}
\label{tab:tunnels}
\end{table}

We have inspected the tunnel surfaces to get information about the hydrophobic
and charged amino-acid residues exposed on the surface from ribosomal proteins.
The numbers of these residues for each tunnel are listed in Table
\ref{tab:tunnels}. The distribution of hydrophobic residues along the 
tunnel axis is shown in Fig.~\ref{fig:4tunnels}(b).
It is found that E. coli has the highest number of hydrophobic residues on the
tunnel surface, about 30\% higher than the other organisms. The hydrophobic
residues are the most abundant near the tunnel exit for E. coli, S. cerevisiae 
and H. sapiens (Fig.~\ref{fig:4tunnels}(b)).
It is interesting to note that almost all the charges of amino acids on the
tunnel surface are positive charges for E. coli, S. cerevisiae and H. sapiens
(Table \ref{tab:tunnels}), suggesting that the charged amino acids play an
important role for the function of the exit tunnel. 
Note that the ribosomal RNA is negatively charged so only positively charged
amino acids can significantly change the electrostatic potential inside the
tunnel.
An exception is found for H. marismortui, for which the number of 
negatively charged amino acids on the tunnel surface is much higher than in the
other organisms, even though it is still significantly smaller than the number
of positively charged ones (19 vs. 26). The distinction of the electrostatics
of H. marismortui's tunnel may be related to the fact that this species 
can survive in extreme environmental conditions, such as at high temperatures,
with high salt concentration, or at high or low pH.

\subsection{Conservation of the diffusion mechanism of the escape process}

We have carried out simulations of the nascent chain's growth and the escape
processes of all proteins considered in the G\=o-MJ model at the four
ribosomal tunnels and in the G\=o-MJ-nn model at the human ribosomal tunnel
only. In most cases, the protein can escape easily at the exit
tunnel, but for several proteins at some of the tunnels, kinetic trapping can
delay the escape. A kinetic trap is found in a simulation if the protein get
stuck in some state at the tunnel leading to a much longer escape
time, more than 10 times longer than in an average trajectory. Kinetic
trapping can be due to the roughness in the shape of the exit tunnel as well as
the interactions between nascent proteins and the tunnel wall
\cite{Thuy2020,Thuy2021}. Interestingly, it has been shown that the probability
of trapping of a protein decreases with the growth time per residue $t_g$, and
can become negligibly small at realistic translation rates \cite{Thuy2021}. In
our study, we have simulated the easily escaped proteins with $t_g=400\tau$,
whereas those with kinetic trapping with the increased $t_g=2000\tau$. The
latter value of $t_g$ reduces the trapping probability to below 5\% and makes
the statistics reliable. We have checked that further increase of $t_g$
produces very little changes to the escape time distribution and the median
escape time.

\begin{figure}
\centering
\includegraphics[width=8.5cm]{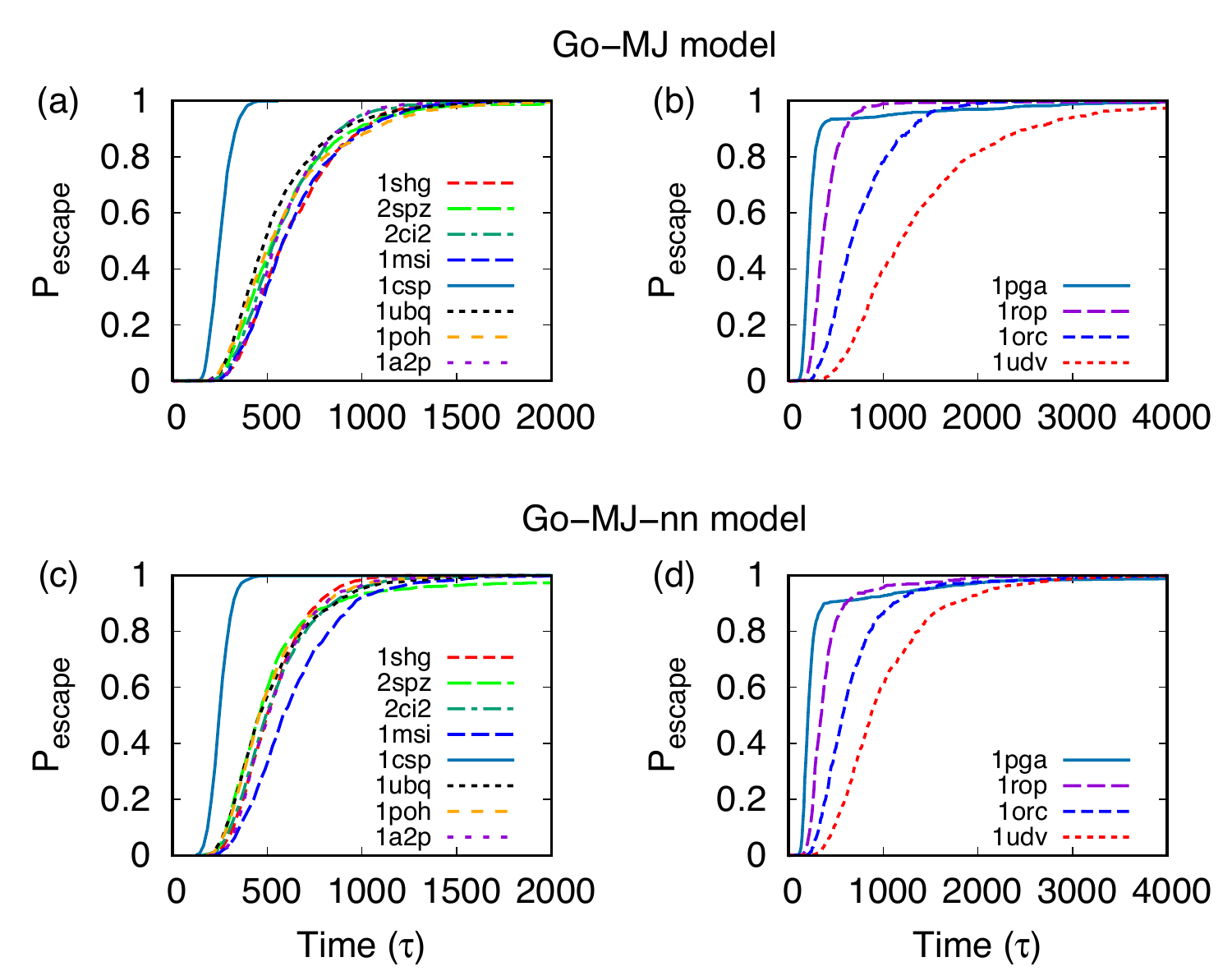}
\caption{The escape probability, $P_{escape}$, as a function of time 
at $T=300$~K for proteins in the G\=o-MJ (a, b) and 
the G\=o-MJ-nn (c, d) model at the human ribosomal exit tunnel.
The 8 proteins in panels (a) and (c) were simulated with $t_g=400\tau$
whereas the 4 proteins in (b) and (d) were simulated with $t_g=2000\tau$. 
An increased value of $t_g$ was used because the latter proteins have 
higher probabilities of kinetic trapping.
}
\label{fig:pesc}
\end{figure}

Figure \ref{fig:pesc} shows that the escape probability, $P_\mathrm{escape}$, 
increases sigmoidally with time and asymptotically approaches the value of 1
for all proteins in both G\=o-MJ and G\=o-MJ-nn models at the human ribosomal
exit tunnel. The proteins also escape efficiently at all other ribosomal exit
tunnels. The proteins that are more likely to get kinetically trapped are 1pga,
1rop, 1orc and 1udv with 1udv being the slowest escaper.  The median escape
time, $t_\mathrm{esc}$, the time at which $P_\mathrm{escape}=0.5$, varies among
the proteins from a few hundred to a few thousand $\tau$ (see Table
\ref{tab:tesc}). 

\begin{figure} 
\centering
\includegraphics[width=8.5cm]{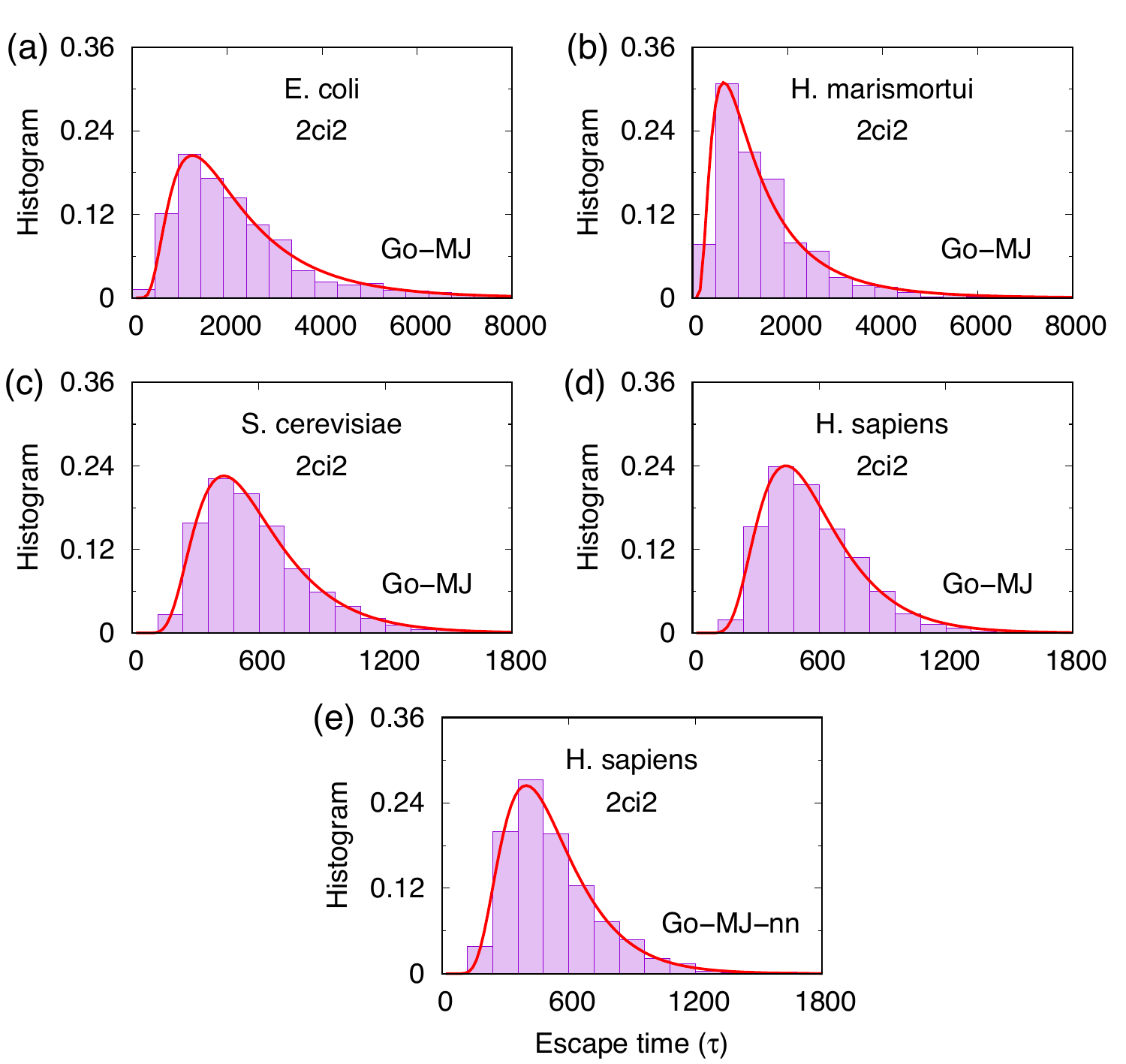}
\caption{Distributions of the escape time for the 2ci2 protein
in the G\=o-MJ model at the ribosomal tunnels of E. coli (a), H. marismortui
(b), S. cerevisiae (c), H. sapiens (d) and in the G\=o-MJ-nn model at
the H. sapiens ribosomal tunnel (e). The normalized histograms obtained by
simulations (boxes) are fitted to the diffusion model (solid line) by using Eq.
(\ref{eq:gt}). The simulations were carried out at the temperature $T
= 300$~K and with the growth time per residue $t_g=400\tau$.}
\label{fig:CI2}
\end{figure}

\begin{table*}
\begin{tabular}{lccrcrcrcrcr}
\hline
 & & \multicolumn{2}{c}{E. coli} &
     \multicolumn{2}{c}{H. marismortui} &
     \multicolumn{2}{c}{S. cerevisiae} &
     \multicolumn{2}{c}{H. sapiens} &
     \multicolumn{2}{c}{H. sapiens} \\
 & & \multicolumn{2}{c}{G\=o-MJ} &
     \multicolumn{2}{c}{G\=o-MJ} &
     \multicolumn{2}{c}{G\=o-MJ} &
     \multicolumn{2}{c}{G\=o-MJ} &
     \multicolumn{2}{c}{G\=o-MJ-nn} \\
Protein & $N$ & 
 $D\,$({\AA}$^2\tau^{-1}$) & $k\,$(pN) & 
 $D\,$({\AA}$^2\tau^{-1}$) & $k\,$(pN) & 
 $D\,$({\AA}$^2\tau^{-1}$) & $k\,$(pN) & 
 $D\,$({\AA}$^2\tau^{-1}$) & $k\,$(pN) & 
 $D\,$({\AA}$^2\tau^{-1}$) & $k\,$(pN) \\ 
\hline
1pga     & 56  & 1.174 & 12.010 & 0.817 & 14.288 & 0.779 & 21.121 & 1.103 & 12.880  & 1.180 & 12.507     \\
1rop     & 56  & 1.093 &  5.135 & 1.030 &  1.822 & 0.771 & 14.495 & 0.891 &  9.028  & 0.911 &  9.484     \\
1shg     & 57  & 0.510 &  6.543 & 0.497 &  4.059 & 0.703 & 11.182 & 0.761 &  6.171  & 0.704 &  8.241     \\
2spz     & 58  & 0.504 & 12.466 & 0.405 & 11.927 & 0.580 & 16.731 & 0.935 &  5.674  & 0.947 &  6.875     \\
1orc     & 64  & 0.797 &  5.881 & 0.338 &  1.905 & 0.754 &  9.484 & 0.877 &  4.597  & 1.008 &  4.763     \\
2ci2     & 65  & 0.487 &  2.650 & 1.065 &  1.822 & 0.909 &  5.591 & 0.802 &  6.461  & 0.873 &  6.585     \\
1msi     & 66  & 0.419 &  6.171 & 0.559 &  3.686 & 0.555 & 14.039 & 0.734 &  6.461  & 0.754 &  6.336     \\
1csp     & 67  & 0.586 & 16.690 & 0.625 & 12.714 & 0.789 & 14.867 & 0.723 & 16.483  & 0.714 & 17.849     \\
1ubq     & 76  & 0.459 & 10.146 & 0.536 &  5.342 & 0.570 & 14.122 & 0.948 &  6.129  & 1.108 &  5.591     \\
1poh     & 85  & 0.712 &  8.532 & 0.571 &  9.235 & 0.962 & 10.271 & 1.148 &  4.597  & 0.954 &  6.543     \\
1udv & 88  & 0.379 &  2.112 & 0.516 &  0.396 & 0.578 &  5.011 & 0.641 &  3.437  & 0.672 &  4.555     \\
1a2p & 108 & 0.710 &  4.141 & 0.780 &  2.650 & 1.064 &  3.396 & 0.610 &  8.490  & 0.671 &  8.738     \\
\hline
\end{tabular}
\caption{Diffusional properties of the protein escape process at 
ribosomal exit tunnels.  The proteins are identified by their PDB ID (first 
column) and
the chain length, $N$. For each protein, the properties given are the 
diffusion constant $D$ and the pulling force $k$ of the diffusion model,
whose values, in units of {\AA}$^2\tau^{-1}$ and pN, respectively,
are obtained by fitting the histograms of escape times from simulations
to the diffusion model (see text). The names of the organisms 
and the model for nascent proteins considered are given on tops of
the $D$ and $k$ columns.
}
\label{tab:dk}
\end{table*}

We inspected the diffusion mechanism of the escape process by examining the
escape time distributions of the proteins. The model mechanism is that of the
diffusion model described in Section II. B, which corresponds to the diffusion
of a one-dimensional Brownian particle in a linear potential field.
Interestingly, for all proteins and all the tunnels considered, the escape time
distribution follows relatively well that of the diffusion model. For example,
Figure \ref{fig:CI2} shows that the histogram of the escape times of the 2ci2
protein obtained by the simulations can be fitted to the distribution function
in Eq. (\ref{eq:gt}) for all the exit tunnels considered. Thus, the
diffusion mechanism is conserved among the proteins and among the species
although the individual distributions can be different from each other.
From the fits to the diffusion model we can get the values of
the parameters $D$ and $k$, which can be considered as an effective diffusion
constant of a protein at a tunnel and an effective mean force acting on the
protein along the escape coordinate, respectively. These are highly collective
quantities which reflect the complex dynamics of nascent proteins at the exit
tunnels. The values of $D$ and $k$ are listed in Table \ref{tab:dk} for all the
proteins in each tunnel. They strongly vary with the protein and with the
tunnel (Fig.~S4). $D$ is in the range from 0.4 to 1.2 {\AA}$^2\,\tau^{-1}$.
With $\tau=3$~ps, the obtained values of $D$ are of the order
of 10$^{-8}$ m$^2$ s$^{-1}$, i.e. about two orders of magnitude larger
than diffusion constants of isolated proteins in water 
($\sim$10$^{-10}$ m$^2$ s$^{-1}$) \cite{Brune1993}. Note that $D$ depends on
the friction coefficient $\zeta$ and the value of $\zeta$ used in the simulations
is 100 times smaller than that of amino acids in water \cite{Thuy2021}.
It is expected that at realistic friction, $D$ is smaller but of the same order
of magnitude to that of isolated proteins.
The force $k$ varies more strongly than $D$. It is interesting that the
obtained values of $k$ are in the range from a sub-piconewton to few tens of
piconewtons, which is within the scale of molecular forces in proteins
\cite{Bustamante2011}. 

\subsection{Conservation of the effects of hydrophobic and electrostatic
interactions on the protein escape time}

\begin{table*} 
\begin{tabular}{lccccccccc}
\hline
 & & 
 & E. coli & H. marismortui & S. cerevisiae & H. sapiens & H. sapiens\\
 & & 
 & G\=o-MJ & G\=o-MJ & G\=o-MJ & G\=o-MJ & G\=o-MJ-nn\\
Protein & $N_h$ & $Q\, (e)$ 
& $t_\mathrm{esc}\, (\tau)$ & $t_\mathrm{esc}\, (\tau)$ 
& $t_\mathrm{esc}\, (\tau)$ & $t_\mathrm{esc}\, (\tau)$ 
& $t_\mathrm{esc}\, (\tau)$ \\
\hline
1pga     & 12 & $-4$ &  220.0 &  248.0 & 177.1 &  204.3 & 195.8 \\ 
1rop     & 15 & $-$4 &  495.6 & 1262.5 & 269.7 &  351.6 & 330.0 \\
1shg     & 20 & $+1$ &  819.0 & 1273.0 & 361.4 &  578.5 & 478.6 \\ 
2spz     & 16 & $-2$ &  463.6 &  595.9 & 295.7 &  523.2 & 436.5 \\
1orc     & 18 & $+$3 &  602.3 & 2587.9 & 391.0 &  649.5 & 564.5 \\
2ci2     & 29 & $-$1 & 1906.0 & 1175.0 & 525.5 &  523.1 & 476.7 \\
1msi     & 31 & 0    & 1050.9 & 1261.7 & 365.2 &  572.4 & 566.5 \\
1csp     & 22 & $-$6 &  294.4 &  359.8 & 244.1 &  240.8 & 226.8 \\
1ubq     & 26 &  0   &  599.0 &  941.4 & 356.5 &  477.1 & 437.9 \\
1poh     & 26 & $-$2 &  463.7 &  532.9 & 287.1 &  507.6 & 436.0 \\
1udv     & 32 & $+$3 & 2828.3 & 5843.2 & 930.0 & 1185.8 & 859.8 \\
1a2p     & 29 & $+$2 &  907.2 & 1202.0 & 727.3 &  536.1 & 485.1 \\
\hline
\end{tabular}
\caption{Median escape times of proteins at the ribosomal exit tunnels of
different organisms. The proteins are listed with the number of hydrophobic
residues, $N_h$, the net charge, $Q$, and the median escape times,
$t_\mathrm{esc}$, given in units of $\tau$ obtained by simulations. The names
of the organism and the protein model used in the simulations are given on top
of each $t_\mathrm{esc}$ column. }
\label{tab:tesc}
\end{table*}

\begin{figure*}
\center
\includegraphics[width=17cm]{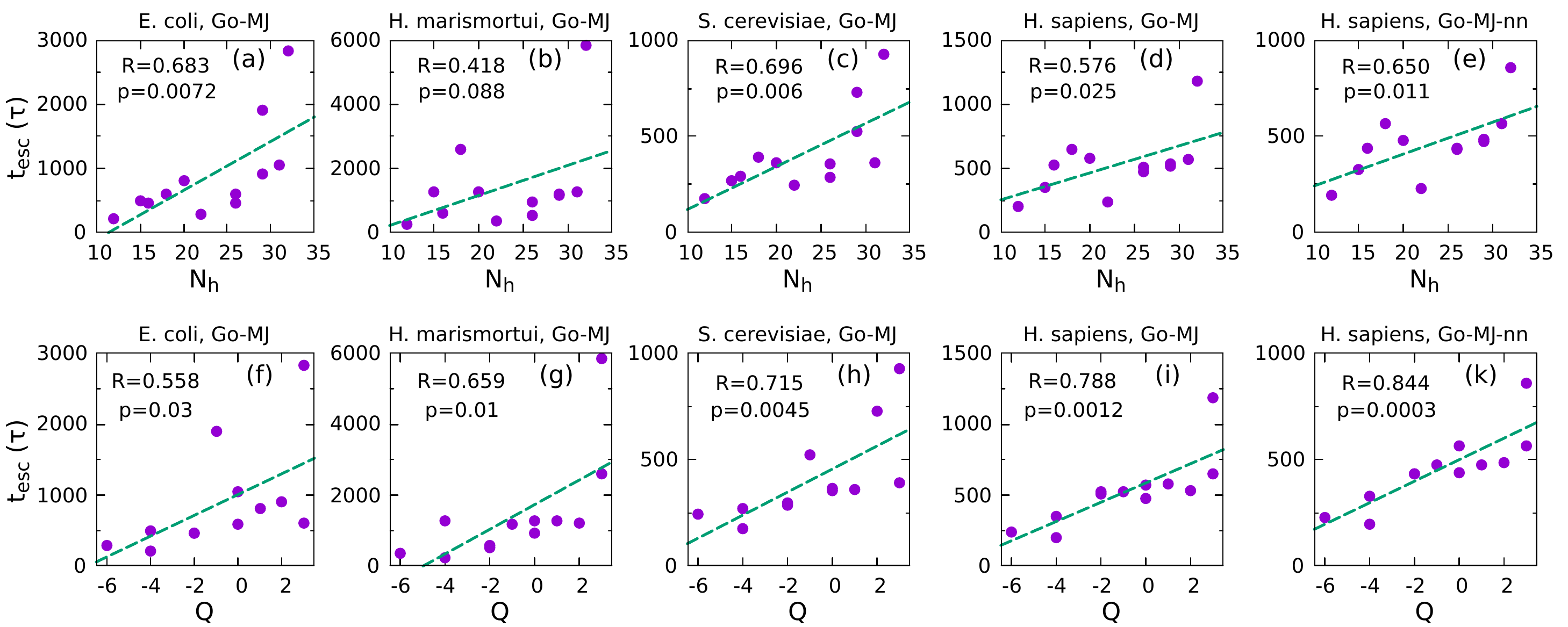}
\caption{Dependence of the median escape time, $t_{esc}$, on the number of
hydrophobic residues, $N_h$, (a--e) and the total charge, $Q$, (f--k) of
nascent proteins at the ribosomal exit tunnels of different species. 
The names
of the species and the protein model are given on top of each panel. Dashed
line represents a linear fit. The Pearson's correlation coefficient $R$ 
and the corresponding $p$-value calculated using the one-tailed Student's
t-test are given in each figure.} \label{fig:corre_all}
\end{figure*}

\begin{figure}
\center
\includegraphics[width=8.5cm]{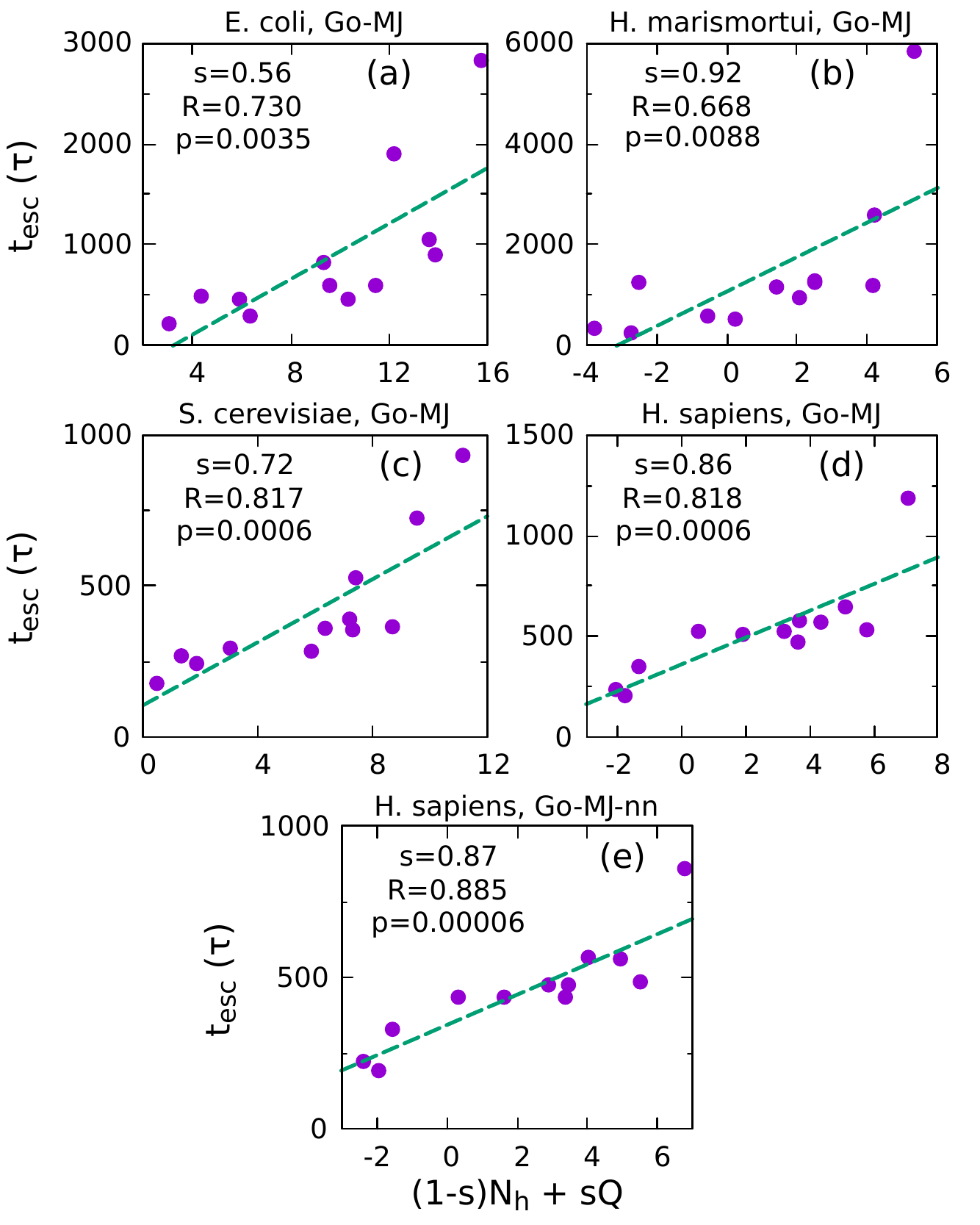}
\caption{Dependence of the median escape time, $t_{esc}$, on the function 
$(1-s)N_h + s Q$ of the number of hydrophobic residues, $N_h$, and 
the total charge, $Q$, of nascent proteins at the ribosomal exit tunnels of
E. coli (a), H. marismortui (b), S. cerevisiae (c), and H. sapiens (d,
e). The proteins are considered in the Go-MJ model (a--d) and 
in the Go-MJ-nn model (e). The names of the species and the protein model are
given on top of each panel. The dependence is shown for the value of $s$ that
maximizes the correlation coefficient $R$.  The values of $s$, $R$,
and $p$ are given
in each panel. Dashed line shows a linear fit of the data.}
\label{fig:corre_rs} \end{figure}

To evaluate the effects of the tunnel's interactions on the protein escape, we
investigated the dependence of the median escape time, $t_{esc}$, on the number
of hydrophobic residues, $N_h$, and the net charge, $Q$, of a protein (Table
\ref{tab:tesc}). Figure \ref{fig:corre_all} shows that $t_{esc}$ positively
correlates with both $N_h$ and $Q$ for all the species considered indicating
that both the hydrophobic and electrostatic interactions modulate the
escape time. The Pearson's correlation coefficient $R$ varies from about
0.42 to about 0.84 for different tunnels and protein models. 
The $p$-values for these correlations, calculated using the one-tailed
Student's t-test, are given in the panels of Fig.~\ref{fig:corre_all}. 
Except the one case shown in Fig.~\ref{fig:corre_all} (b) for the H.
marismortui's tunnel, which shows a weak correlation between $t_{esc}$
and $N_h$ ($R=0.418$ and $p=0.088$), all other correlations have medium to
high $R$-values and are statistically significant ($p < 0.05$). 
The correlation of $t_{esc}$ with $Q$ is higher than the correlation with $N_h$
for all the tunnels except for E. coli. Thus, the effect of electrostatic
interaction on the protein escape tend to be stronger than that of the
hydrophobic interaction, though this still depends on the tunnel.

It is expected that the C-terminal segment of a protein is  
most relevant to its escape process \cite{Nissley2020}.  We have checked that
by calculating $N_h$ and $Q$ only for the C-terminal 50 residues, as shown in
Fig.~S9, the correlation of $t_{esc}$ with $N_h$ becomes very poor or almost
disappears for all the tunnels, while the correlation with $Q$ remains
statistically significant and even slightly improves for some tunnels compared
to the cases without the C-terminal cut-off. This result indicates that the
impact of
electrostatic interaction on the protein escape time is dominating over
hydrophobic interaction for residues near the protein C-terminus.
The impact of hydrophobic interaction on the protein escape seems to appear
with a longer protein segment (the longest protein in our study is barnase with
108 residues), as indicated by the correlations in Fig.~\ref{fig:corre_all}.
However, in longer proteins, it is unlikely that residues too distant from the
C-terminus can influence the escape time.

Following the previous work \cite{Thuy2021}, we tested the dependence of
$t_{esc}$ on a linear function of both $N_h$ and $Q$. The form of the function
chosen is $(1-s)N_h + s Q$ where $s \in (0,1)$ is a tunable parameter. 
We find that this function yields a better correlation with $t_{esc}$ than
both $N_h$ and $Q$ alone at some intermediate value of $s$ (Fig.~S6). 
Figure \ref{fig:corre_rs} plots the dependence of $t_{esc}$ on the function
$(1-s)N_h + sQ$ for the optimal value of $s$, i.e. the value that maximizes
the correlation coefficient $R$, for all the tunnels and the protein models
considered. The values of $R$ in these plots range from 0.668 to 0.885
and all the $p$-values are $< 0.01$. 
The best correlations are found for the tunnels of S. cerevisiae and H. sapiens
with $R$ exceeding 0.8. For the H. sapiens tunnel, the G\=o-MJ-nn model yields
a better correlation than the G\=o-MJ model.

The above results show that the effects of hydrophobic and electrostatic 
interactions on the protein escape time at different exit tunnels are
qualitatively similar. Generally, increasing $N_h$ and $Q$ leads to an
increased escape time, though the quantitative effects depends on the
protein and the tunnel. 

The common mechanism of these effects is that attraction between the protein
and the tunnel slows down the protein escape while repulsion speeds it up
\cite{Thuy2021}. The slowest escape is found for protein 1udv at the H.
marismortui's tunnel with $t_{esc} \approx 5800\tau$ (Fig.~S5) due to
a strong electrostatic attraction between the protein and the tunnel. The 1udv
protein
has the highest net charge among the proteins ($Q=+3e$) (Table \ref{tab:tesc})
while the H.  marismortui's tunnel has the highest number of negatively charged
amino
acid residues on the tunnel's surface ($N_{-}^{(t)}=19$) (Table
\ref{tab:tunnels}). We have checked that switching off the electrostatic
interaction of 1udv drastically reduces its median escape time by 10 times
to about $600\tau$. An example for a strong effect of hydrophobic interaction
is of the protein 2ci2 at the E. coli's tunnel with $t_{esc} \approx 1900\tau$.
The 2ci2 is
among the proteins with the highest numbers of hydrophobic residues ($N_h=29$)
(Table \ref{tab:tesc}) and E. coli has the highest number of hydrophobic residues on the
tunnel's surface ($N_h^{(t)}=46$) (Table \ref{tab:tunnels}).

\subsection{Effects of non-native interactions on the escape process}

The effects of non-native interactions on the protein escape time can be seen
by comparing the results of the G\=o-MJ and the G\=o-MJ-nn models at the 
human ribosomal exit tunnel. It is shown that the two models yield
qualitatively and quantitatively similar results, in the escape time
distribution (Fig.  \ref{fig:CI2}) as well as in the dependence of of $t_{esc}$
on $N_h$ and $Q$ (Figs. \ref{fig:corre_all} and \ref{fig:corre_rs}). A more
careful examination shows that non-native interactions reduce the escape time
by 4\% to 28\% depending on the protein (Table \ref{tab:tesc}). This reduction
effect is consistent with a result
for homopolymer models, which shows that self-attractive polymers escape faster
than self-repulsive polymers for polymer lengths larger than about 60 residues
(Fig.~S7). Non-native interactions that are governed by attractive potentials
energetically drive the protein escape because the chain can form more
non-native contacts if it is found outside the tunnel. 
Even though the native contacts can be more competitive in energy compared to
the non-native contacts as assumed in the G\=o-MJ-nn model, they are fewer in
numbers and some of the native contacts cannot be formed once the protein has
not been fully escaped. This effect can be seen for protein 2ci2
by looking at the distributions of native and non-native contacts, and
radius of gyration of the protein conformations at the moment of complete 
translation at the human ribosomal tunnel (Fig.~S8). 
These distributions show that, on average, the G\=o-MJ-nn model yields
more compact conformations with smaller numbers of native contacts and 
larger numbers of non-native contacts than the G\=o-MJ model.

\section{Discussion}

It is suggested that the driving forces for the protein escape come
from different sources, including (i) an enthalpic preference associated with
the folding of a nascent protein near the tunnel
\cite{Bustamante2015,Thuy2016}, (ii) an entropy gain of a chain
emerging from the tunnel
\cite{OBrien2018}, and (iii) the stochastic motion of a partially folded chain
leading to a kind of diffusion process \cite{Thuy2016,Thuy2018}. The
mechanical forces of the first two types, 
in ribosome stalling and ribosome bound systems,
have been quantified experimentally
\cite{Bustamante2015,OBrien2018} and also
computationally \cite{OBrien2018,OBrien2019}, giving values from several to
about 12 piconewtons. Apart from these sources, the force governing
the protein escape may also come from interactions of nascent chains with the
ribosomal tunnel.  Electrostatic repulsion and attraction as well as
hydrophobic attraction can speed up or slow down the escape process
\cite{Nissley2020,Thuy2021}. 

The mean force from all the above sources acting on a nascent protein at
the ribosome may be well represented by the force $k$ in the diffusion
model, which can be obtained by non-equilibrium methods via the escape time
distribution.  Interestingly, the magnitude of $k$ found in the present
study for different proteins and ribosomes (Table~\ref{tab:dk}) is similar to
the mechanical forces reported in other studies
\cite{Bustamante2015,OBrien2018,OBrien2019}.
For a tunnel that has energetic interactions with nascent proteins, the
force acting on a protein by the tunnel may contribute or counter balance the
other forces depending on whether it is attractive or repulsive.  If the
attraction to the tunnel is sufficiently strong, the protein may have very long
escape times and may not follow the diffusion model. The slowest escaping
protein in our consideration is 1udv at the H. marismortui's model, having a
very small force $k=0.396$ pN. The escape time distribution of this protein has
a long tail but still follows the diffusion model (Fig. S5).  Note that the
diffusion model predicts that for $k=0$ both the mean escape time and the
dispersion diverge (see Methods). It is possible that some proteins can have $k
\leq 0$ resulting in infinite escape times. 

In a recent work \cite{Nissley2020}, Nissley et al. have reported very long
ejection times of some proteins at the E. coli's tunnel, including the ones
with PDB IDs 2jo6, 1u0b and 4dcm, using similar coarse-grained simulations. We
have checked that the fractions of hydrophobic amino acids as well as the
fractions of positively and negatively
charged amino acids of these three proteins are within the ranges given by the
12 proteins considered. Furthermore, 2jo6 and 4u0b have negative net charges
for the whole chain (for the C-terminal 50 residues, the net charge is
zero for 2jo6 and $-1e$ for 1u0b),
suggesting that they are not slow escapers due to overall
electrostatic repulsion with the tunnel. Indeed, our simulations of 2jo6 and
1u0b at the E. coli's tunnel, using an averaged value of $\epsilon$ from
the 12 proteins studied, show that they escape efficiently with the median
escape time $t_{esc} \approx 666\tau$ for 2jo6 and $t_{esc} \approx 1270\tau$
for 1u0b (Fig.  S10), i.e. within the same range of escape times as for other
proteins. It is also found that
their escape time distributions are consistent with
the conserved mechanism given by the diffusion model (Fig.~S10).
Including these two proteins into the initial set of 12 proteins, however,
slightly deteriorates the correlations of $t_{esc}$ with $N_h$ and $Q$
(Fig.~S11). It would be interesting to check what specific detail causes 
the extreme delay of the escape process in Nissley et al.'s approach.
We did not simulate 4dcm because the PDB structure of this protein is not
contiguous containing missing residues. This protein, however, is expected to
escape very slowly because it has a positive net charge of $Q=+8e$ in the 
C-terminal 50 residues, compared to $+5e$ in 1udv, the slowest escaping protein
in the 12 proteins considered.

The previous work \cite{Thuy2021} has estimated the timescale of the protein
escape times to be of the order of 0.1--1 ms by rescaling the simulation
times to the values at the realistic friction and simultaneously using the
high-friction value of the time unit, $\tau_H = 3$~ns
\cite{Veitshans1997,Klimov1997}. In accordance to this estimation, 
the longest median escape times of the proteins in the present study 
are of the order of 10 ms, which is still shorter than the times needed
by the ribosome to translate one codon. Given that the 12 proteins
considered can be representative for most proteins in the proteomes in terms of
hydrophobic and charge fractions as discussed in Section III.A, this result
suggests that typical proteins escape efficiently at the ribosome tunnel and do
not delay the ribosome's new translation cycle.

\section{Conclusion}

The ribosomal exit tunnel has many structural and chemical elements that
could affect the post-translational escape of nascent proteins. These 
elements include the irregular shape of the tunnel, the exposed hydrophobic
side-chains of the ribosomal proteins and the charged amino acids on tunnel's
surface. The exit tunnels from different organisms of different domains of
life, as the ones considered in our study, show significant differences in
structural and physico-chemical properties beside certain similarities. The
present study shows that despite all these differences of the exit tunnels, the
protein escape process has conserved mechanisms across the domains of life.
First, it is shown that the escape process follows the simple diffusion
mechanism described by the diffusion model. This property holds true for twelve
proteins of distinct native structures and diverse physico-chemical
properties within two different protein models that are given with and without
non-native interactions. Second, the median escape time, $t_{esc}$, positively
correlates with both the number of hydrophobic residues, $N_h$, and the net
charge, $Q$, of protein, with a sufficient statistical significance in most
cases. This property underlines the simple mechanism that attraction between
the protein and the tunnel slows down the protein escape while repulsion speeds
it up. The effects of hydrophobic and electrostatic interactions on the escape
time are additive to each other as indicated by improved correlations when
considering the dependence of $t_{esc}$ on a linear function of $N_h$ and $Q$.
These results reinforce our understanding of the protein escape process as the
one that is simple and predictable. 
The results also suggest that the impact of electrostatic interaction
on the escape time is stronger than that of hydrophobic interaction and becomes
dominant towards the C-terminal residues of nascent proteins. It is
expected that proteins with a high positive net charge in the C-terminal
segment may have unusually long escape times. 

Our study also shows significant variations among the organisms when 
considering the quantitative effects of the exit tunnels on the escape process.
The exit tunnels of E. coli and H. marismortui generally yield longer 
protein escape times than that of S. cerevisiae and H. sapiens. These
observations are related to the facts that E. coli has $\sim$30\% higher number
of hydrophobic residues exposed inside the exit tunnel than the other
organisms, and H.  marismortui has the number of negatively charged amino acids
on the tunnel's surface that is substantially larger than the
other organisms (19 vs. 1 to 4). From an evolutionary perspective, it could be
that the exit tunnels of S. cerevisiae and H. sapiens have evolved to deal with a
larger number of proteins in their genomes in suppressing the escape time. The
argument for this hypothesis is that a too slow escape of a nascent protein
from the exit tunnel could hamper the ribosome productivity, thus it is
beneficial to have an exit tunnel that allows all proteins in the genome to
escape efficiently.

\section*{Supplementary Material}

See supplementary material for the list of 12 proteins considered with
selected properties, for the structural class compositions of Richardson's
Top2018 dataset and our protein set,
for the distributions of protein length in various proteomes,
for the histograms of hydrophobicity and charges in protein sequences
of various proteomes,
for the dependences
of the diffusion constant $D$ and the force $k$ from the diffusion
model on the chain length, $N$, of proteins, for the escape properties of
protein 1udv, for the dependence of the correlation coefficient $R$
between the median escape time and the function $(1-s)N_h + sQ$ on the
parameter $s$ for the tunnels and protein models considered, for the escape
properties of self-repulsive and self-attractive homopolymers,
for an analysis of the effects of non-native interactions on the escaping
conformations, for the correlations of $t_{esc}$ with $N_h$ and $Q$ 
calculated for the C-terminal 50 residues,
for the histograms of escape times and the escape probabilities of 2jo6 and 1u0b
proteins, and for the correlations of $t_{esc}$ with $N_h$ and $Q$ by adding
2jo6 and 1u0b to the initial protein set.

\section*{Acknowledgements}
This research is funded by Graduate University of Science and Technology under
grant number GUST.STS.DT2020-VL03 for the postdoctoral fellowship of Phuong
Thuy Bui. We also acknowledge the supports of the Institute of Physics
and the Centre for Informatics and Computing of VAST for
the use of their computer clusters.

\section*{Data availability statement}

The data that support the findings of this study are available from the
corresponding author upon reasonable request.

\bibliography{refs_tunnel5}

\newpage

\setcounter{equation}{0}
\renewcommand\theequation{S\arabic{equation}}

\setcounter{figure}{0}
\renewcommand\thefigure{S\arabic{figure}}

\setcounter{table}{0}
\renewcommand\thetable{S\arabic{table}}

\setcounter{page}{1}
\renewcommand{\bibnumfmt}[1]{[S#1]}
\renewcommand{\citenumfont}[1]{[S#1]}

\onecolumngrid

\makeatletter

\centerline{\large\bf Supplementary Material}

\begin{center}
{\large\bf
The protein escape process at the ribosomal exit tunnel has conserved
mechanisms across the domains of life}
\end{center}

\centerline{P. T. Bui and T. X. Hoang}

\vspace{20pt}

\begin{table*}[!ht] 
\begin{tabular}{lcccccccccccc}
\hline
 & & & & & & & & &  
\multicolumn{2}{c}{G\=o-MJ model} & \multicolumn{2}{c}{G\=o-MJ-nn model} \\
Protein name & ID & Class & $T_m\,$($^\circ$C) & 
$N$ & $N_h$ & $N_{+}$ & $N_{-}$ & $Q\,(e)$ &
$T_\mathrm{max}\,(\epsilon/k_B)$ & $\epsilon\,$(kcal/mol) 
& $T_\mathrm{max}\,(\epsilon/k_B)$ & $\epsilon\,$(kcal/mol) 
\\
\hline
Protein G, B1 domain & 1pga & $\alpha/\beta$  &   87.5 & 56 & 
12 & 6 & 10 & $-4$ &
0.9298 & 0.770475 & 0.9522 & 0.752349 \\
Rop protein & 1rop & $\alpha$ &  68.7 &  56 & 
15 & 7 & 11 & $-$4 &
0.7921 & 0.857250 & 0.7435 & 0.913285 \\
SH3 domain & 1shg & $\beta$  &  82.0 & 57 & 
20 & 10 & 9 & $+1$ &
0.9977 & 0.707084 & 1.0330 & 0.682954 \\
Protein A, Z domain & 2spz & $\alpha$ &  78.0 & 58 & 
16 & 7 & 9 & $-2$ &
0.8906 & 0.783190 & 0.8759 & 0.796334 \\
Cro repressor & 1orc & $\alpha/\beta$ & 57.0 &  64 & 
18 & 10 & 7 & $+$3 &
0.8747 & 0.749717 & 0.8232 & 0.796620 \\
CI2 & 2ci2 & $\alpha/\beta$  &  80.0 & 65 & 
29 & 10 & 11 & $-$1 &
0.9375 & 0.748249 & 0.9445 & 0.742703 \\
Antifreeze protein & 1msi & $\beta$ & 46.6 &  66 & 
31 & 4 & 4 & 0 &
1.0019 & 0.633906 & 0.9834 & 0.645831 \\
Cold-shock protein & 1csp & $\alpha/\beta$ & 43.8 & 67 & 
22 & 6 & 12 & $-$6 &
0.9368 & 0.672018 & 0.9712 & 0.648215 \\
Ubiquitin & 1ubq & $\alpha/\beta$  &  95.0 & 76 & 
26 & 11 & 11 & 0 &
0.9718 & 0.752512 & 0.9959 & 0.734302 \\
HPr protein & 1poh & $\alpha/\beta$ & 63.4 &  85 & 
26 & 8 & 10 & $-$2 &
0.9570 & 0.698532 & 0.9152 & 0.730437 \\
Sso10b2 & 1udv & $\alpha/\beta$ & 157.5 &  88 & 
32 & 16 & 13 & $+$3 &
0.9921 & 0.862304 & 0.9422 & 0.907972 \\
Barnase & 1a2p & $\alpha/\beta$ & 55.0 & 108 & 
29 & 14 & 12 & $+$2 &
0.9795 & 0.665445 & 0.9881 & 0.659653 \\
\hline
\end{tabular}
\caption{List of 12 proteins considered with selected properties.  Each protein
is given with the PDB ID of its native
structure (ID), the structure classification (Class), the experimental melting
temperature ($T_m$), the protein length in the number of amino acid residues
($N$), the number of hydrophobic residues ($N_h$), the numbers of positively
($N_+$) and negatively ($N_-$) charged residues, and the total charge ($Q$).
Also shown are the temperature of the specific heat maximum ($T_\mathrm{max}$)
and the energy parameter ($\epsilon$) obtained for each protein in the G\=o-MJ
and G\=o-MJ-nn models. 
}
\label{properties}
\end{table*}

\begin{table*}[!ht]
\begin{tabular}{ccccccc}
\hline
      &\ \ \ & \multicolumn{2}{c}{Richardson's Top2018 dataset}
      &\ \ \ & \multicolumn{2}{c}{Present study's protein set}  \\
Structural Class & & Number of proteins\ & \%
& & Number of proteins\ & \% \\
\hline
$\alpha$ & &  1,164  & 8.6\%    & & 2      & 16.7\% \\
$\beta$  & &  261  & 1.9\%    & & 2      & 16.7\% \\
$\alpha/\beta$ & & 12,257 & 89.6\% & & 8   & 66.6\% \\
Total    & & 13,677  & 100\%    & & 12     & 100\% \\
\hline
\end{tabular}
\caption{Number and percentage of proteins from different structural classes
in the Richardson's Top2018 dataset of high-quality protein structures 
[C. J. Williams, D. C. Richardson, \&  J. S. Richardson, Prot. Sci. 31, 290–300
(2022)] at a 70\% structural homology level and in the protein set of the
present study. }
\label{sclass}
\end{table*}

\begin{figure}
\center
\includegraphics[width=10cm]{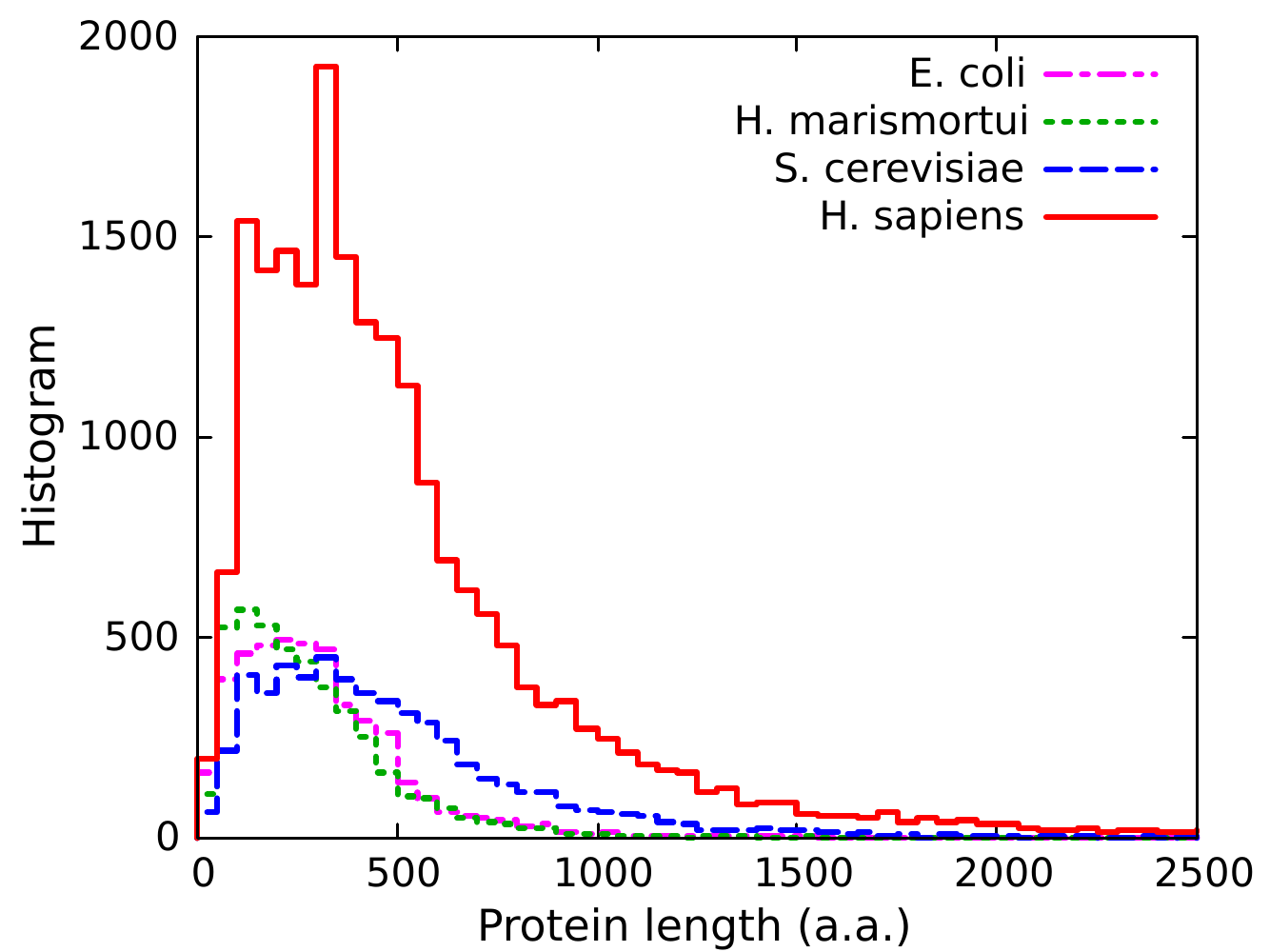}
\caption{Distributions of protein length (in number of amino acids) in the
proteomes of E. coli (dot-dashed), H. marismortui (dotted), S. cerevisiae
(dashed) and H. sapiens (solid). The protein sequences of the proteomes are
taken from the UniProt database [The UniProt Consortium, Nucl. Acid. Res. 49,
D480 (2021)] with one sequence per gene.}
\end{figure}

\begin{figure}
\center
\includegraphics[width=14cm]{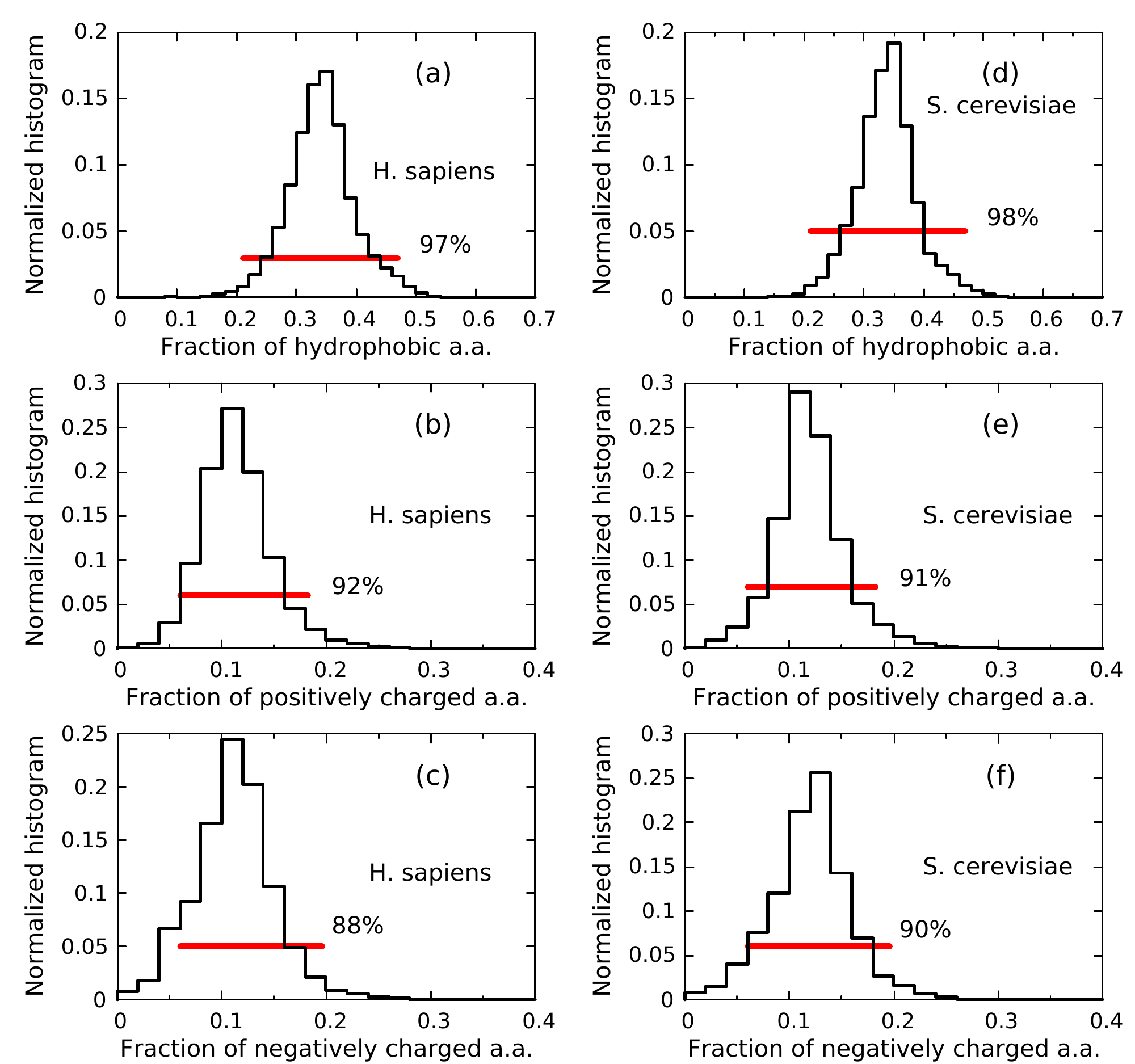}
\caption{Histograms of the fraction of hydrophobic amino acids (a, d), the
fraction of positively charged amino acids (b, e), and the fraction of
negatively charged amino acids (c, f) of proteins in the proteomes of H.
sapiens (a--c) and S. cerevisiae (d--f). The human proteome (ID:
UP000005640) contains 20,607 protein sequences and the S. cerevisiae's one
(ID: UP000470054) contains 5,551 sequences with one protein sequence
per gene.  The proteome data are obtained from the UniProt
database. Horizontal bars (red) indicate the ranges of the above fractions in
the 12 proteins
considered in the present study, as listed in Table S1.  The ranges associated
with the horizontal bars correspond to $\sim$97\%, $\sim$92\% and $\sim$88\% of
the protein population for the histograms in (a), (b) and (c), respectively,
and correspond to $\sim$98\%, $\sim$91\% and $\sim$90\% of the protein
population for the histograms in (d), (e) and (f), respectively.  }
\end{figure}

\begin{figure}
\center
\includegraphics[width=14cm]{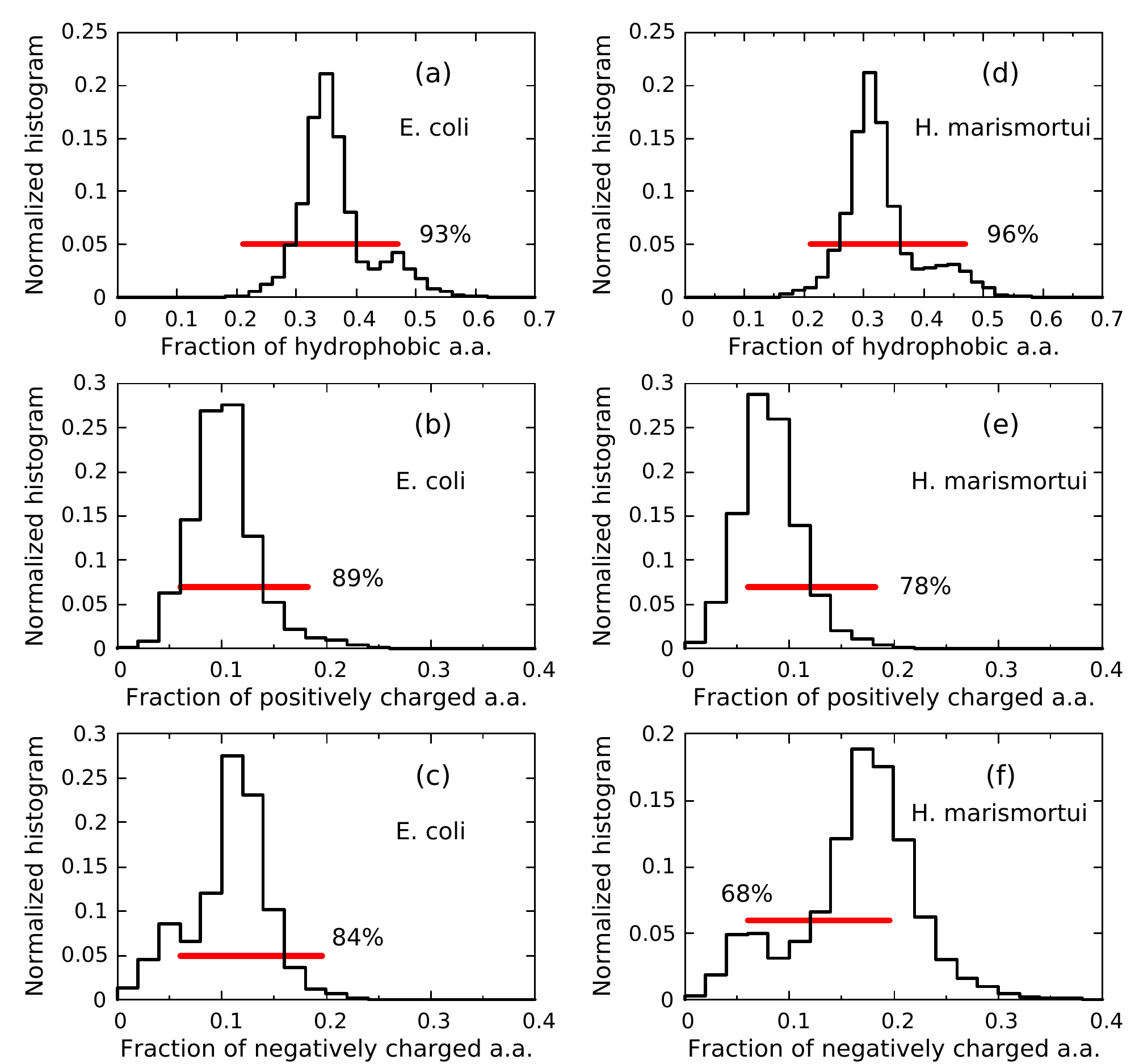}
\caption{Histograms of the fraction of hydrophobic amino acids (a, d), the
fraction of positively charged amino acids (b, e), and the fraction of
negatively charged amino acids (c, f) of proteins in the proteomes of E. coli
(a--c) and H. marismortui (d--f). The E. coli's proteome (ID: UP000000625)
contains 4,402 protein sequences and the H. marismortui's one
(ID: UP000001169) contains 4,234 sequences with one
protein sequence per gene.  The proteome data are obtained from the UniProt
database. Horizontal bars (red) indicate the ranges of the above fractions in
the 12 proteins
considered in the present study, as listed in Table S1.  The ranges associated
with the horizontal bars correspond to $\sim$93\%, $\sim$89\% and $\sim$84\% of
the protein population for the histograms in (a), (b) and (c), respectively,
and correspond to $\sim$96\%, $\sim$78\% and $\sim$68\% of the protein
population for the histograms in (d), (e) and (f), respectively.  }
\end{figure}

\begin{figure}[!ht]
\center
\includegraphics[width=8.5cm]{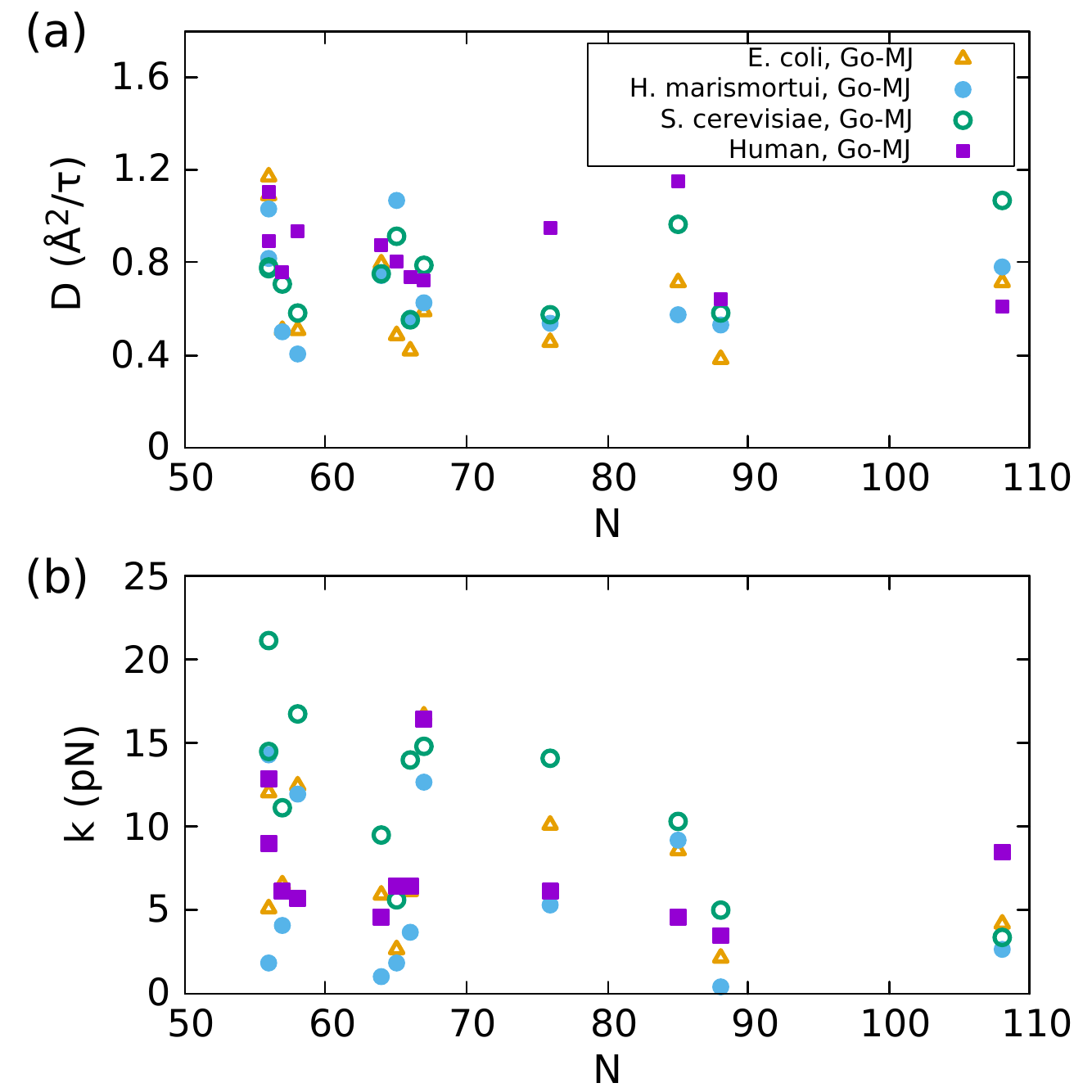}
\caption{Dependence of the diffusion constant $D$ and the pulling force $k$ from
the diffusion model on the chain length, $N$, of proteins. The data are shown
for the escape processes of proteins in the G\=o-MJ model at the exit tunnels
of E. coli (open triangles), H. marismortui (filled circles), S. cerevisiae
(open circles) and H. sapiens (filled squares). The data for the G\=o-MJ-nn model
in the H. sapiens' exit tunnel are not shown because they are very close to that
of the G\=o-MJ model. The values of $D$ and $k$ are
listed in Table 2 of the main text.}
\label{}
\end{figure}

\begin{figure}[!ht]
\center
\includegraphics[width=12cm]{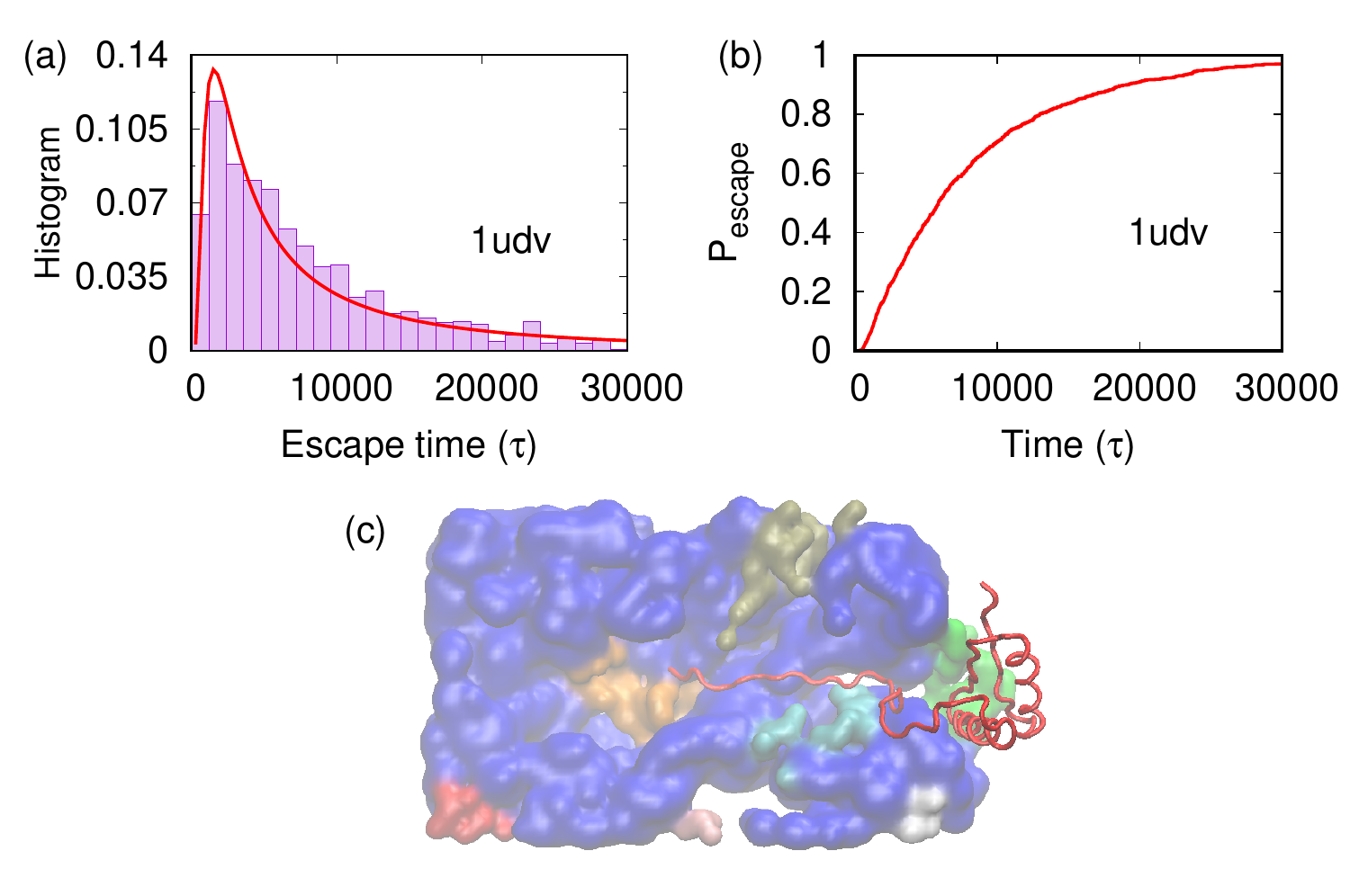}
\caption{The escape time distribution (a) and the escape probability as
a function of time (b) of protein Sso10b2 (1udv) at the H. marismortui's tunnel.
(c) A slowly escaping conformation of 1udv at the tunnel. This protein escapes
slowly mainly due to electrostatic interactions between the protein 
and the tunnel.}
\label{fig:udv}
\end{figure}

\begin{figure}[!ht]
\center
\includegraphics[width=8.5cm]{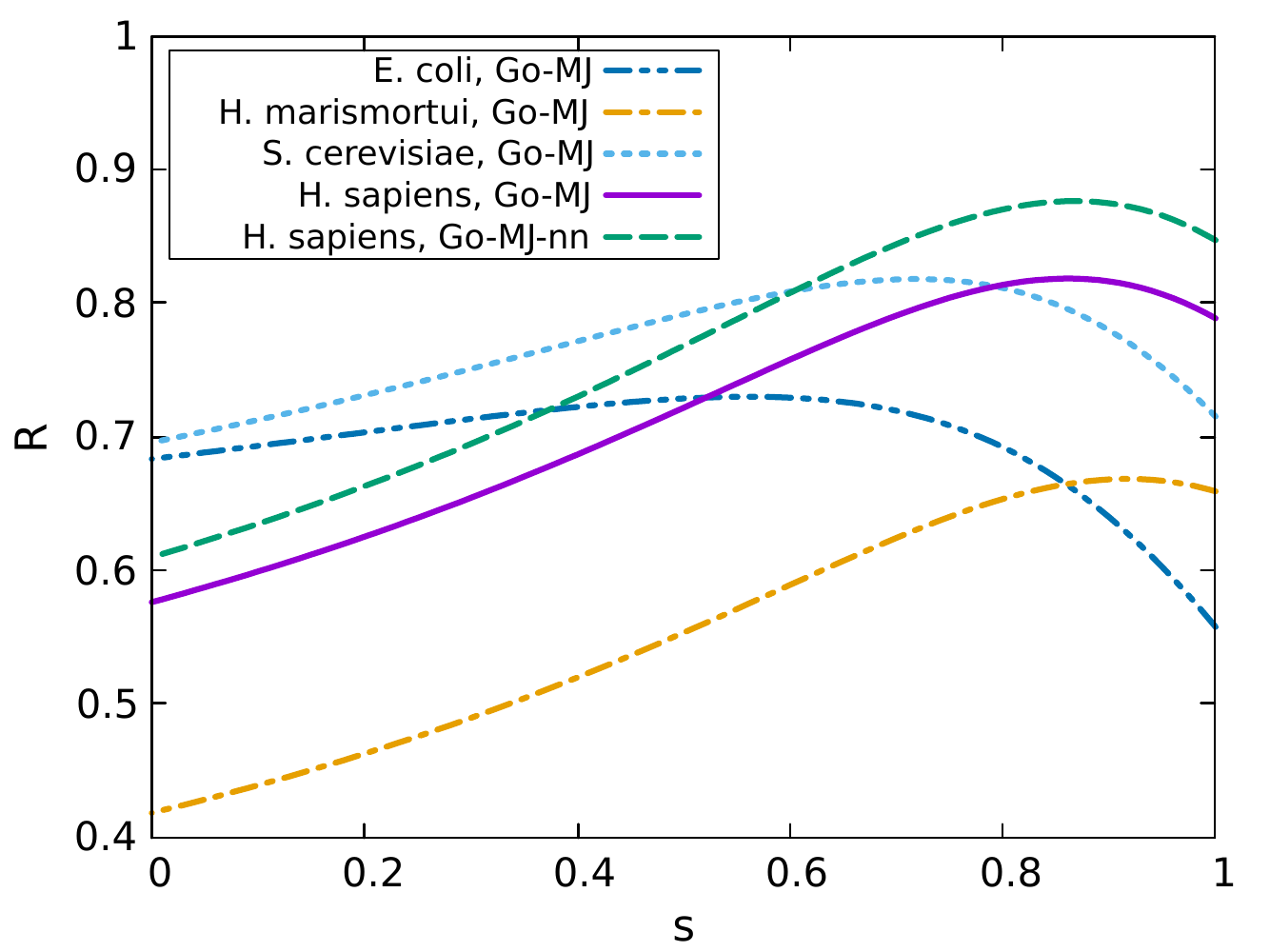}
\caption{Dependence of the correlation coefficient $R$ between the median
escape time, $t_\mathrm{esc}$, and the function $(1-s) N_h + s Q$, of the
proteins considered in the present study (Table S1)
on the parameter $s$ for different species and protein models as indicated.}
\label{fig:rapha}
\end{figure}

\begin{figure}
\center
\includegraphics[width=12cm]{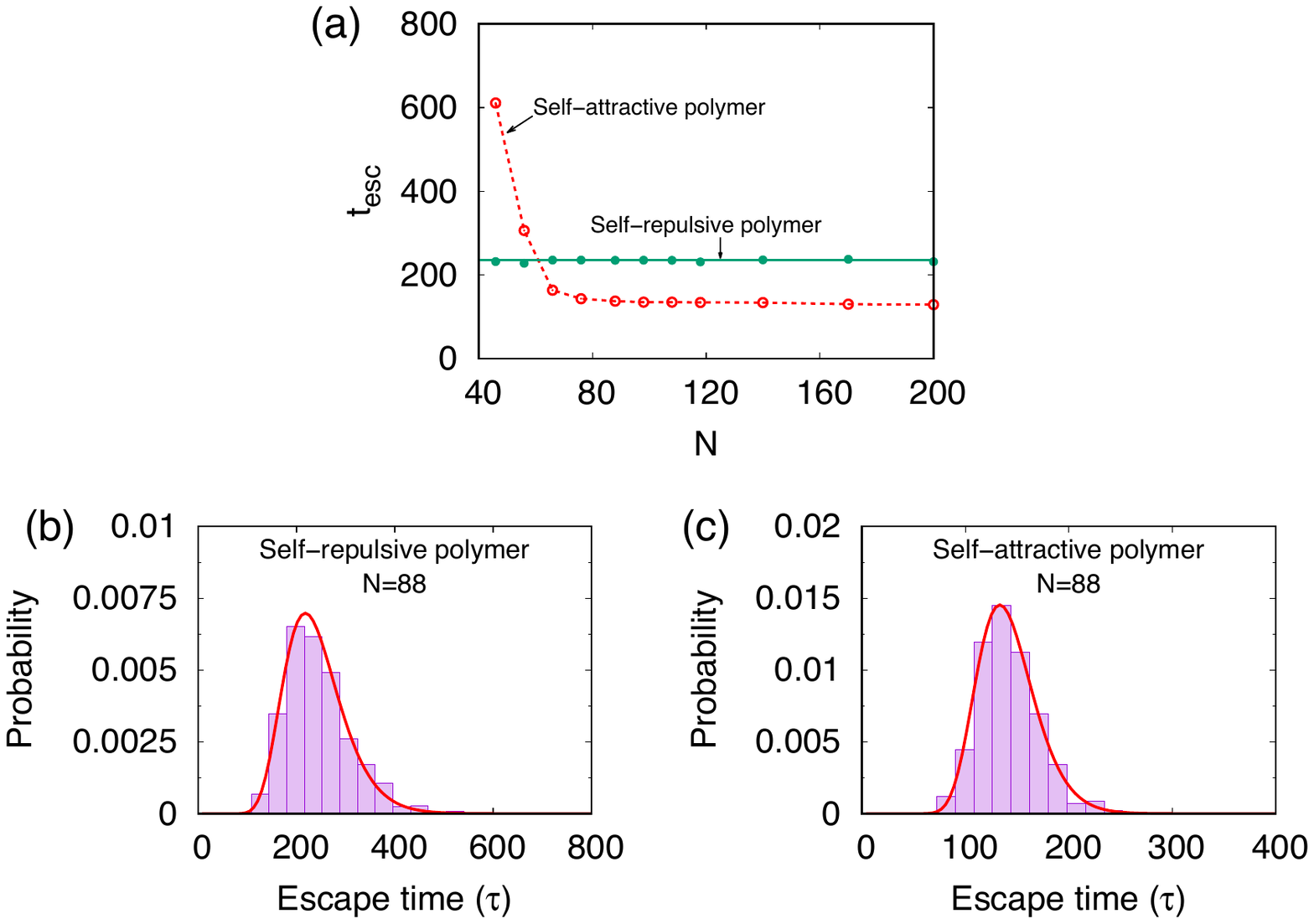}
\caption{(a) Dependence of the median escape time,
$t_{esc}$, on the chain length, $N$, of the self-repulsive (filled circles) and
self-attractive (open circles) homopolymers at the human ribosomal tunnel. 
A self-repulsive homopolymer is modelled as a chain with a repulsive
potential of $\epsilon (\sigma/r)^{12}$ for the interaction between any pair of
non-consecutive beads. A self-attractive one employs
the 12-10 LJ potential, as given by Eq. (3) of the main text, but with the
potential depth given by $\epsilon$, for the attraction
between the beads. The value of $\epsilon=0.75$~kcal/mol is used
for both types of the homopolymers. The polymer chains are grown inside the
tunnel with $t_g=400\tau$ in the simulations as for nascent proteins.
The data were obtained at the temperature $T = 300K$ for the human tunnel.
(b and c) The escape time distributions for a $N=88$ self-repulsive homopolymer
(b) and a $N=88$ self-attractive homopolymer (c). The histograms obtained
by simulations are fitted to the diffusion model (solid line).
} \label{polymer_t_esc_ss}
\end{figure}

\begin{figure}
\center
\includegraphics[width=12cm]{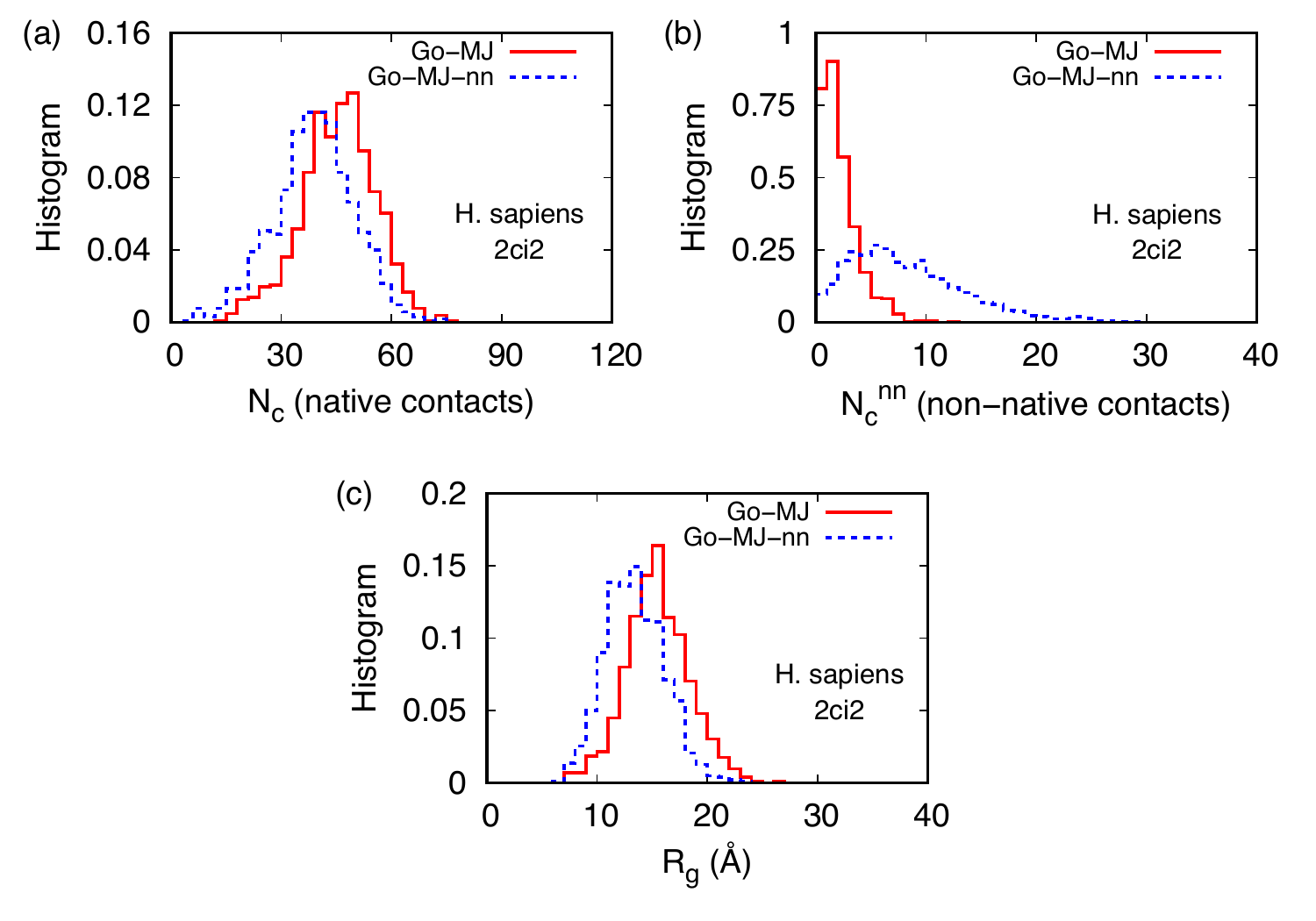}
\caption{Normalized histograms of number of native contacts $N_c$ (a), 
number of non-native contacts $N_c^{nn}$ (b), and the radius of gyration $R_g$ (c)
of protein 2ci2 at the moment of complete translation 
at the human ribosomal tunnel. The statistics are drawn from 1000 conformations
obtained by independent trajectories in the Go-MJ model (solid) and the
Go-MJ-nn model (dotted). The contacts are defined based on a cut-off distance
of $1.2\, r_{ij}^*$ for native contacts and $1.2\,\sigma_1$ for non-native
contacts (see Methods).
}
\label{fig:ncnn}
\end{figure}

\begin{figure}[!ht]
\center
\includegraphics[width=17cm]{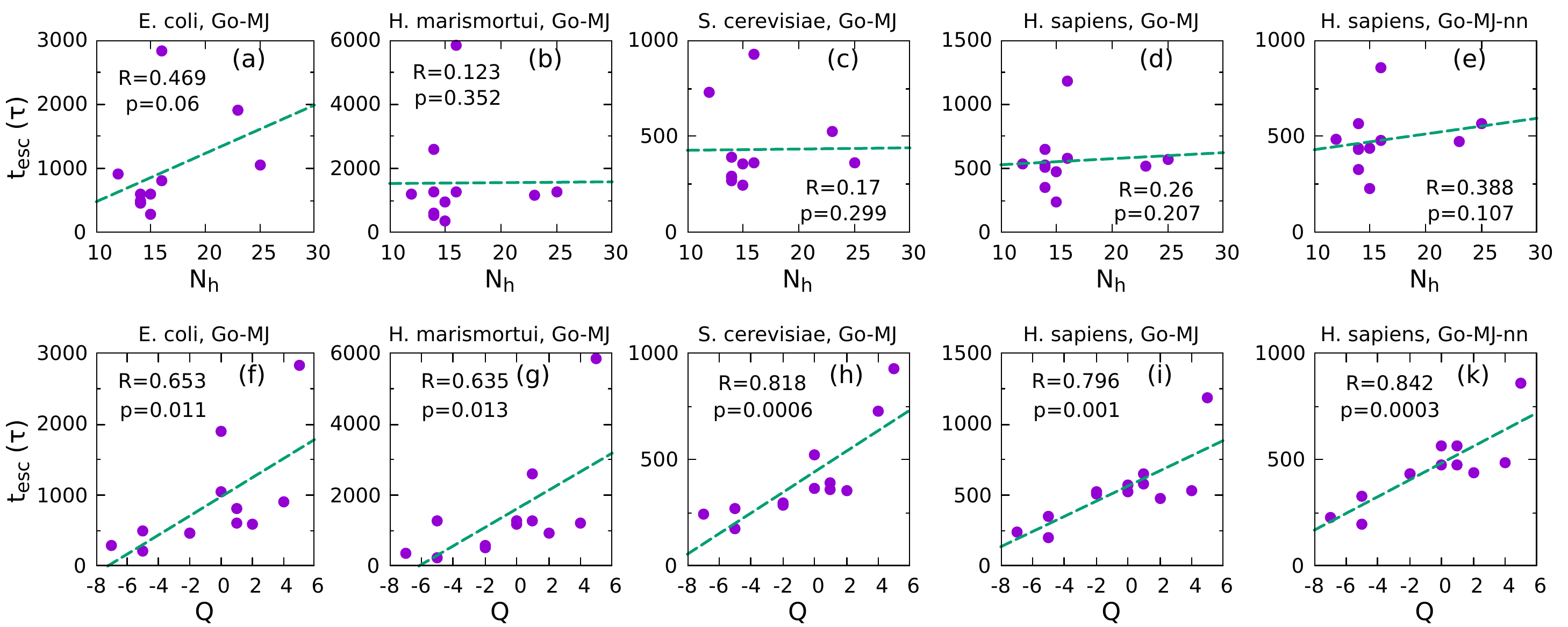}
\caption{Correlations between the median escape time $t_{esc}$ obtained
by simulations and the number of hydrophobic residues $N_h$ (a--e), and
between $t_{esc}$ and the net charge $Q$ (f--k) of the proteins considered
for the organisms and the protein models, as indicated.
Here, $N_h$ and $Q$ are obtained only for the C-terminal 50 residues.
The correlation coefficient $R$
and the corresponding $p$-value are given in each panel.}
\label{fig:corre60}
\end{figure}

\begin{figure}
\center
\includegraphics[width=11cm]{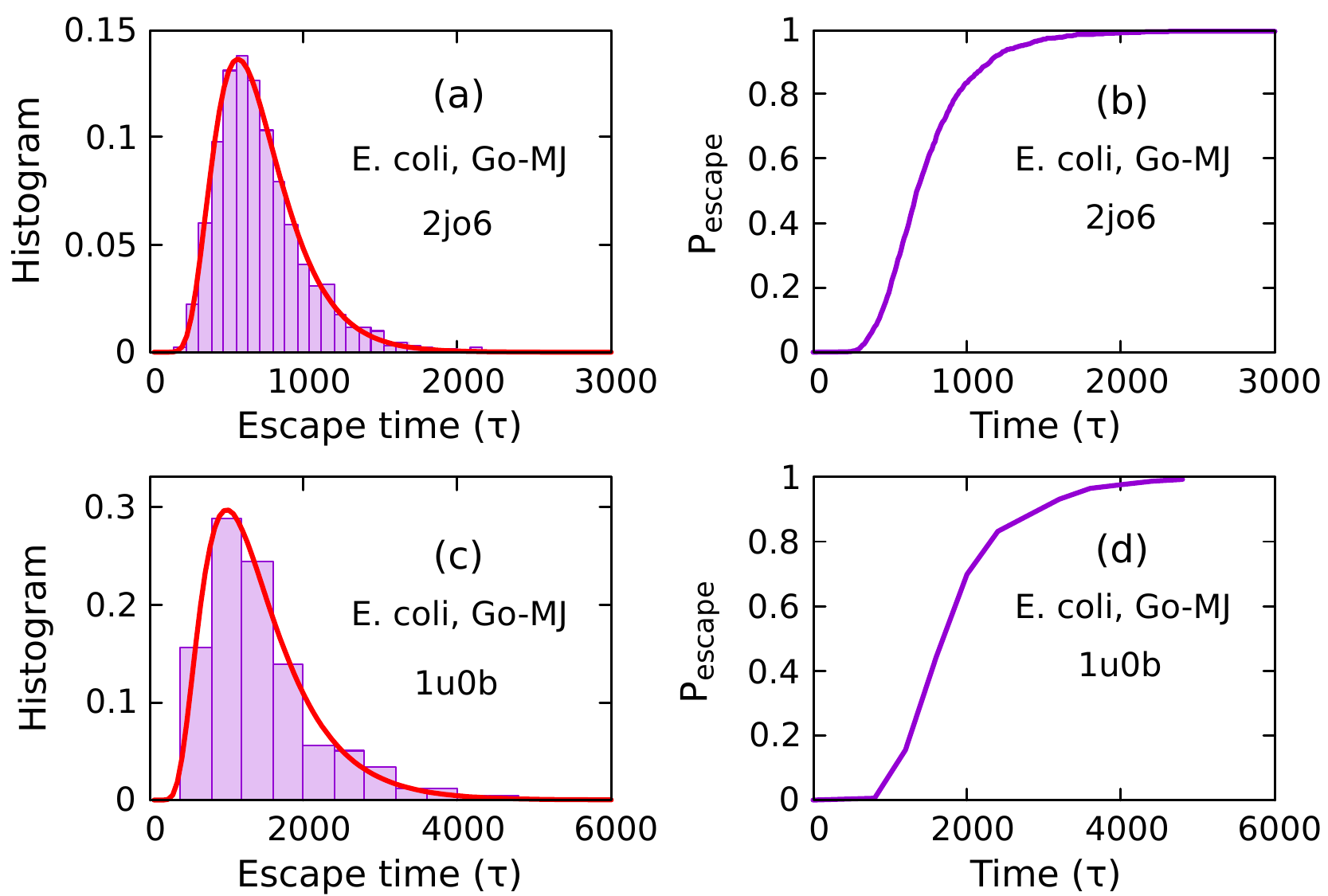}
\caption{Histogram of the escape time (a, c) and the time dependence of
the escape probability $P_\mathrm{escape}$ (b, d) of proteins 2jo6 and 1u0b,
which are identified by their PDB IDs, at the ribosomal exit tunnels of E. coli.
The proteins are considered in the Go-MJ model with $\epsilon=0.7417$~kcal/mol,
which is the mean value of $\epsilon$ for the 12 proteins listed in
Table S1. The 2jo6 protein has the length of $N=110$ residues, in which
the number of hydrophobic residues is $N_h=40$, the numbers of positively and
negatively charged residues are $N_{+}=12$ and $N_{-}=17$, respectively, giving
the net charge $Q=-5e$.
The 1u0b protein has the length of $N=461$ residues with $N_h=148$, $N_{+}=53$,
$N_{-}=72$, giving $Q=-19e$.
For C-terminal 50 residues, the net charge is zero for 2jo6 and $-1e$ for
1u0b.
The histogram shown are obtained from simulations with 1000 independent
trajectories for 2jo6 and 180 independent trajectories for 1u0b with
$t_g=400\tau$. The median escape times found are $t_{esc} \approx 666\tau$ for
2jo6 and $t_{esc}\approx 1270\tau$ for 1u0b.
} \label{fig:protnew}
\end{figure}

\begin{figure}
\center
\includegraphics[width=11cm]{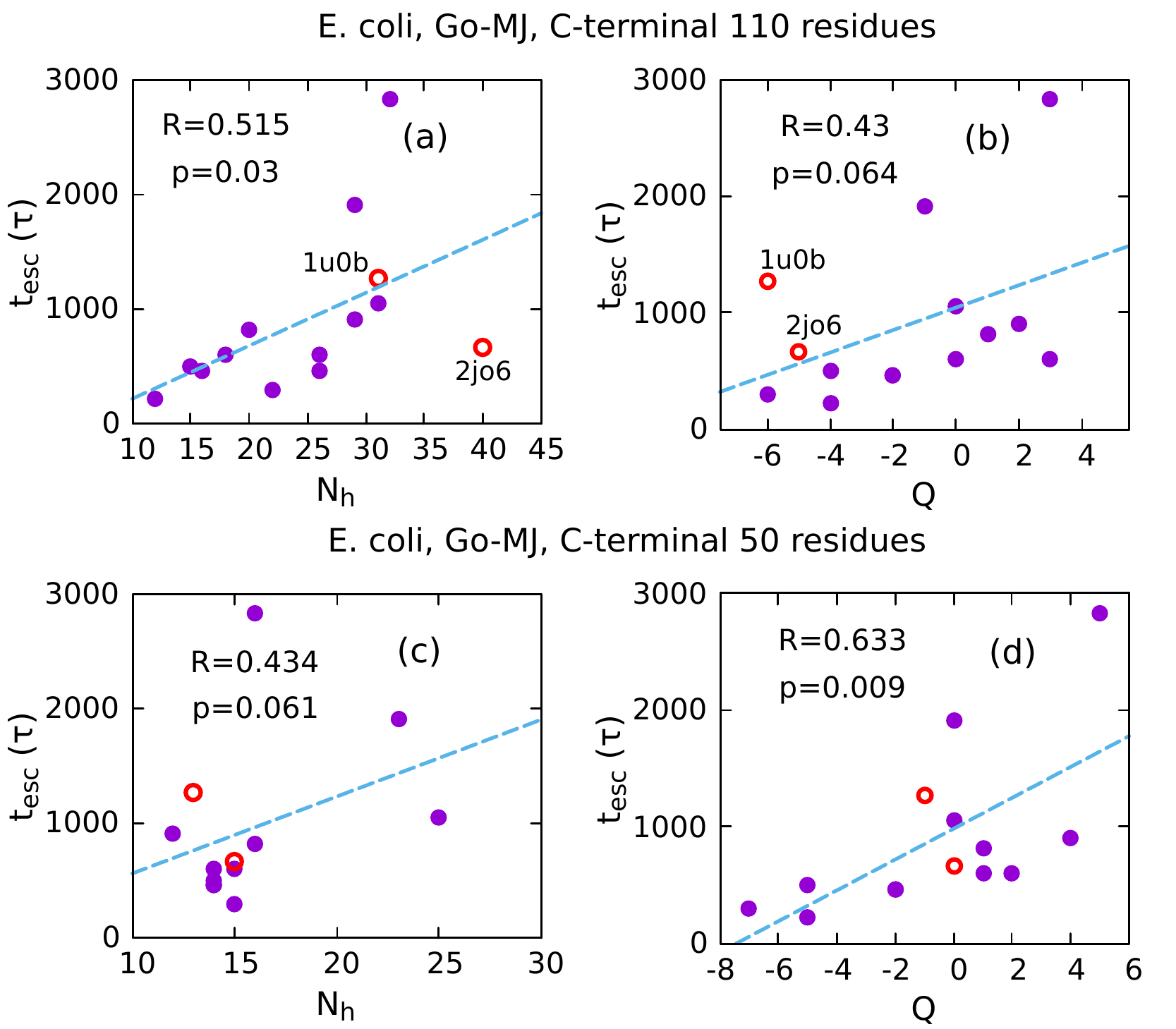}
\caption{Correlations of the median escape time $t_{esc}$ 
with the number of hydrophobic residues $N_h$ (a, c) and the net charge
$Q$ (b, d) of nascent proteins at the E. coli's exit tunnel. The data shown
include those of the 12 proteins considered initially (closed circles), as
listed in Table S1, and two new proteins, 2jo6 and 1u0b, (open circles).
Top panels (a, b) correspond to $N_h$ and $Q$ obtained for 
for C-terminal 110 residues if applicable.  Bottom panels
(c, d) correspond to $N_h$ and $Q$ obtained for C-terminal 50 residues.
$R$-values and $p$-values for the correlations are given in each panel. }
\label{fig:corrnew}
\end{figure}

\end{document}